\documentclass[a4paper,12pt]{article}

\pdfoutput=1
\usepackage{amsmath,amssymb}
\usepackage{graphicx}
\usepackage{subcaption}
\usepackage[numbers,sort&compress]{natbib}
\usepackage{xcolor}
\usepackage[colorlinks,
            linkcolor=black,
            filecolor=black,
            anchorcolor=black,
            urlcolor=black,
            citecolor=blue
            ]{hyperref}
\usepackage{ulem}
\usepackage{setspace}
\usepackage{array}
\usepackage{booktabs}

\newlength{\dinwidth}
\newlength{\dinmargin}
\newcommand{\SM}{{\rm SM}}
\newcommand{\NP}{{\rm NP}}

\newcommand{\GeV}{{\,\rm GeV}}
\newcommand{\TeV}{{\,\rm TeV}}

\newcommand{\ab}{{\, \rm ab}}

\newcommand{\mB}{{\mathcal B}}

\newcommand{\mC}{{\mathcal C}}

\newcommand{\bt}{{\tilde{b}}}

\newcommand{\lp}[1]{\lambda_{3 #1 3}^\prime}
\newcommand{\lps}[1]{\lambda_{3 #1 3}^{\prime *}}

\begin{document}

\setlength{\abovedisplayskip}{6pt}
\setlength{\belowdisplayskip}{6pt}

\title{\bf{$\boldsymbol{b\rightarrow c\tau\bar{\nu}}$ decays in supersymmetry with $\boldsymbol{R}$-parity violation}}

\author{
Dong-Yang Wang\footnote{wangdongyang@mails.ccnu.edu.cn},\,
Ya-Dong Yang\footnote{yangyd@mail.ccnu.edu.cn},\,
and
Xing-Bo Yuan\footnote{y@mail.ccnu.edu.cn}\\[15pt]
\small Institute of Particle Physics and Key Laboratory of Quark and Lepton Physics~(MOE), \\
\small Central China Normal University, Wuhan, Hubei 430079, China}
\date{}
\maketitle

\vspace{1em}

\begin{abstract}

  {\noindent}In the past few years, several hints of lepton flavour universality (LFU) violation have emerged in the $b \to c \tau \bar\nu$ and $b \to s \ell^+ \ell^-$ data. Quite recently, the Belle Collaboration has reported the first measurement of the $D^*$ longitudinal polarization fraction in the $B \to D^* \tau \bar\nu$ decay. Motivated by this intriguing result, together with the recent measurements of $R_{J/\psi}$ and $\tau$ polarization, we study $b \to c \tau \bar\nu$ decays in the Supersymmetry (SUSY) with $R$-parity violation (RPV). We consider $B \to D^{(*)} \tau \bar\nu$, $B_c \to \eta_c \tau \bar\nu$, $B_c \to J/\psi \tau \bar\nu$ and $\Lambda_b \to \Lambda_c \tau \bar\nu$ modes and focus on the branching ratios, the LFU ratios, the forward-backward asymmetries, polarizations of daughter hadrons and $\tau$ lepton. It is found that the RPV SUSY can explain the $R_{D^{(*)}}$ anomalies at $2\sigma$ level, after taking into account various flavour constraints. In the allowed parameter space, the differential branching fractions and LFU ratios are largely enhanced by the SUSY effects, especially in the large dilepton invariant mass region. In addition, a lower bound $\mB(B^+ \to K^+ \nu \bar\nu) > 7.37 \times 10^{-6}$ is obtained. These observables could provide testable signatures at the High-Luminosity LHC and SuperKEKB, and correlate with direct searches for SUSY.

\end{abstract}

\newpage

\section{Introduction}
\label{sec:intro}

In recent years, several interesting anomalies emerge in experimental data of semi-leptonic $B$-meson decays. For the ratios $R_{D^{(*)}} \equiv \mathcal B(B\to D^{(*)}\tau\bar\nu) / \mathcal B(B\to D^{(*)}\ell\bar\nu)$ with $\ell = e,\mu$, the latest averages of the measurements by BaBar~\cite{Lees:2012xj,Lees:2013uzd}, Belle~\cite{Huschle:2015rga,Sato:2016svk,Hirose:2016wfn,Hirose:2017dxl} and LHCb Collaboration~\cite{Aaij:2015yra,Aaij:2017uff,Aaij:2017deq} give~\cite{HFLAV}
\begin{align}
R_D^{\rm exp}&=0.407\pm 0.039 \, ({\rm stat.}) \pm 0.024 \, ({\rm syst.}),  \\\nonumber
R_{D^*}^{\rm exp}&=0.306\pm 0.013 \, ({\rm stat.})\pm 0.007 \, ({\rm syst.}) .
\end{align}
Compared to the branching fractions themselves, these ratios have the virtue that, apart from significant reduction of the experimental systematic uncertainties, the CKM matrix element $V_{cb}$ cancels out and the sensitivity to $B\to D^{(\ast)}$ transition form factors becomes much weaker. The SM predictions read~\cite{HFLAV}
\begin{align}
  R_D^\SM&=0.299\pm 0.003,\\\nonumber
  R_{D^*}^\SM&=0.258\pm 0.005,
\end{align}
which are obtained from the arithmetic averages of the most recent calculations by several groups~\cite{Bigi:2016mdz,Jaiswal:2017rve,Bernlochner:2017jka,Bigi:2017jbd}. The SM predictions for $R_D$ and $R_{D^*}$ are below the experimental measurements by $2.3\sigma$ and $3.0\sigma$, respectively. Taking into account the measurement correlation of $-0.203$ between $R_D$ and $R_{D^*}$, the combined experimental results show about $3.78\sigma$ deviation from the SM predictions~\cite{HFLAV}.  For the $B_c \to J/\psi \tau \bar\nu$ decay, which is mediated by the same quark-level process as $B \to D^{(*)} \tau \bar\nu $, the recent measured ratio $R_{J/\psi}^{\rm exp}=0.71\pm 0.17\,({\rm stat.}) \pm 0.18\, ({\rm syst.})$ at the LHCb~\cite{Aaij:2017tyk} lies within about $2\sigma$ above the SM prediction $R_{J/\psi}^\SM=0.248 \pm 0.006$~\cite{Wang:2008xt}. In addition, the LHCb measurements of the ratios $R_{K^{(*)}}\equiv \mB(B \to K^{(*)} \mu^+ \mu^-)/\mB(B \to K^{(*)} e^+ e^-)$, $R_K^{\rm exp}=0.745_{-0.074}^{+0.090}\pm0.036$ for $q^2\in [1.0, 6.0] \GeV^2$~\cite{Aaij:2014ora} and $R_{K^*}^{\rm exp}=0.69_{-0.07}^{+0.11}\pm 0.05$ for $q^2 \in [1.1,6.0]\GeV^2$~\cite{Aaij:2017vbb}, are found to be about $2.6\sigma$ and $2.5\sigma$ lower than the SM expectation, $R_{K^{(*)}}^{\rm SM}\simeq1$~\cite{Hiller:2003js,Bordone:2016gaq}, respectively. These measurements, referred to as the $R_{D^{(*)}}$, $R_{J/\psi}$ and $R_{K^{(*)}}$ anomalies, may provide hints of Lepton Flavour University (LFU) violation and have motivated numerous studies of New Physics (NP) both in the Effective Field Theory (EFT) approach~\cite{Sakaki:2014sea,Bhattacharya:2014wla,Calibbi:2015kma,Alonso:2015sja,Alonso:2016gym,Feruglio:2016gvd,Ligeti:2016npd,Bardhan:2016uhr,Dutta:2016eml,Bhattacharya:2016zcw,Bordone:2017anc,Choudhury:2017qyt,Bhattacharya:2018kig,Hu:2018veh} and in specific NP models~\cite{Sakaki:2012ft,Crivellin:2012ye,Fan:2015kna,Kim:2015zla,Dorsner:2016wpm,Dumont:2016xpj,Hiller:2016kry,Faroughy:2016osc,Bhattacharya:2016mcc,Wang:2016ggf,Popov:2016fzr,Celis:2016azn,Wei:2017ago,Cvetic:2017gkt,Ko:2017lzd,Chen:2017eby,Crivellin:2017zlb,Cai:2017wry,Iguro:2017ysu,DiLuzio:2017vat,Calibbi:2017qbu,He:2017bft,Fuyuto:2017sys,Li:2018rax,Angelescu:2018tyl,Kim:2018oih}. We refer to refs.~\cite{Li:2018lxi,Bifani:2018zmi} for recent reviews.

Recently, the first measurement on the $D^*$ longitudinal polarization fraction in the $B \to D^* \tau \bar\nu$ decay has been reported by the Belle Collaboration~\cite{Adamczyk:2019wyt,Abdesselam:2019wbt}
\begin{align*}
  P_L^{D^*}=0.60 \pm 0.08 \, (\text{stat.}) \pm 0.04 \, (\text{syst.}),
\end{align*}
which is consistent with the SM prediction $P_{L}^{D^*}=0.46 \pm 0.04$~\cite{Alok:2016qyh} at $1.5\sigma$. Previously, the Belle Collaboration also performed measurements on $\tau$ polarization in the $B \to D^* \tau \bar\nu$ decay, which gives the result $P_L^\tau =-0.38 \pm 0.51 \, (\text{stat.})_{-0.16}^{+0.21} \, (\text{syst.})$~\cite{Hirose:2016wfn,Hirose:2017dxl}. Angular distributions can provide valuable information about the spin structure of the interaction in the $B \to D^{(*)}\tau \bar\nu$ decays, and are good observables to test various NP explanations~\cite{Tanaka:2010se,Tanaka:2012nw,Huang:2018nnq,Iguro:2018vqb,Asadi:2018sym}. Measurements of the angular distributions are expected to be significantly improved in the future. For example, Belle II with $50\,{\rm ab}^{-1}$ data can measure $P_L^\tau$ with a precision of $\pm 0.07$~\cite{Kou:2018nap}.

In this work, motivated by these recent experimental progresses, we study the $R_{D^{(*)}}$ anomalies in the Supersymmetry (SUSY) with $R$-parity violation (RPV). In this scenario, the down-type squarks interact with quarks and leptons via the RPV couplings. Therefore, they contribute to $b \to c \tau \bar\nu$ transition at the tree level and could explain the current $R_{D^{(*)}}$ anomalies~\cite{Deshpande:2012rr,Deshpand:2016cpw,Altmannshofer:2017poe}. Besides $B \to D^{(*)} \tau \bar\nu$, we will also study $B_c \to J/\psi \tau  \bar\nu$, $B_c \to \eta_c \tau \bar\nu$, and $\Lambda_b \to \Lambda_c \tau \bar\nu$ decay. All of them are $b \to c \tau \bar\nu$ transition at the quark level, and the latter two decays have not been measured yet. Using the latest experimental data of various low-energy flavour processes, we will derive constraints on the RPV couplings. Then, predictions in the RPV SUSY are made for the five $b \to c \tau \bar\nu$ decays, focusing on the $q^2$ distributions of the branching fractions, the LFU ratios and various angular observables. We have also taken into account the recent developments on the form factors~\cite{Wang:2008xt,Detmold:2015aaa,Bigi:2016mdz,Bigi:2017jbd,Datta:2017aue}. Implications for future searches at the High-Luminosity LHC (HL-LHC) and SuperKEKB are briefly discussed.

This paper is organized as follows: In section~\ref{sec:SUSY}, we briefly review the SUSY with RPV interactions. In section~\ref{sec:observables}, we recapitulate the theoretical formulae for the various flavour processes, and discuss the SUSY effects. In section~\ref{sec:Numerical results}, detailed numerical results and discussions are presented. We conclude in section~\ref{sec:conclusions}. The relevant form factors are recapitulated in appendix~\ref{sec:form factor}.

\section{Supersymmetry with $\boldsymbol{R}$-parity violation}
\label{sec:SUSY}

The most general renormalizable RPV terms in the superpotential are given by~\cite{Barbier:2004ez,Chemtob:2004xr}
\begin{align}
W_{RPV} = \mu_i L_i H_u + \frac{1}{2} \lambda_{ijk}L_i L_j E^c_k + \lambda^\prime_{ijk}L_i Q_j D^c_k + \frac{1}{2} \lambda^{\prime\prime}_{ijk}U^c_iD^c_jD^c_k,
\end{align}
where $L$ and $Q$ denote the $SU(2)$ doublet lepton and quark superfields, respectively. $E$ and $U$ ($D$) are the singlet lepton and quark superfields, respectively. $i$, $j$ and $k$ indicate generation indices. In order to ensure the proton stability, we assume the couplings $\lambda_{ijk}^{\prime\prime}$ are zero. In semi-leptonic $B$ meson decays, contribution from the $\lambda$ term is through the exchange of sleptons and much more suppressed than the one from the $\lambda^\prime$ term, which is through the exchange of right-handed down-type squarks~\cite{Deshpande:2012rr}. Therefore, we only consider the $\lambda^\prime_{ijk}L_iQ_jD_k^c$ term in this work. For the SUSY scenario with the $\lambda$ term, studies on the $R_{D^{(*)}}$ anomalies with slepton exchanges can be found in ref.~\cite{Zhu:2016xdg,Wei:2018vmk}.

The interaction with $\lambda^\prime_{ijk}$ couplings can be expanded in terms of fermions and sfermions as~\cite{Deshpande:2012rr}
\begin{align}
  \Delta\mathcal{L}_{\rm RPV} &= -\lambda'_{ijk} \Big[\tilde{\nu}_L^i \bar{d}_R^k d_L^j + \tilde{d}_L^j \bar{d}_R^k {\nu}_L^i + \tilde{d}_R^{k*} \bar{{\nu}}_R^{ci} d_L^j \nonumber
  \\
                              & \hspace{3em} -  V_{jl } \big ( \tilde{\ell}_L^i \bar{d}_R^k u_L^l + \tilde{u}_L^l \bar{d}_R^k \ell_L^i + \tilde{d}_R^{k*} \bar{\ell}_R^{ci} u_L^l \big ) \Big]+ {\rm h.c.},
\end{align}
where $V_{ij}$ denotes the CKM matrix element. Here, all the SM fermions $d_{L,R}$, $\ell_{L,R}$ and $\nu_L$ are in their mass eigenstate. Since we neglect the tiny neutrino masses, the PMNS matrix is not needed for the lepton sector. For the sfermions, we assume that they are in the mass eigenstate.  We refer to ref.~\cite{Barbier:2004ez} for more details about the choice of basis. Finally, we adopt the assumption in ref.~\cite{Altmannshofer:2017poe} that only the third family is effectively supersymmetrized. This case is equivalent to that the first two generations are decoupled from the low-energy spectrum as in ref.~\cite{Brust:2011tb,Papucci:2011wy}. For the studies including the first two generation sfermions, we refer to ref.~\cite{Deshpand:2016cpw}, where both the $R_{D^{(*)}}$ and $R_{K^{(*)}}$ anomalies are discussed.

It is noted that the down-type squarks and the scalar leptoquark (LQ) discussed in ref.~\cite{Bauer:2015knc} have similar interaction with the SM fermions. However, in the most general case, the LQ can couple to the right-handed $SU(2)_L$ singlets, which is forbidden in the RPV SUSY. Such right-handed couplings are important to explain the $(g-2)_\mu$ anomaly in the LQ scenario~\cite{Bauer:2015knc}. These couplings can also affect semi-leptonic $B$ decays. In particular, their contributions to the $B \to D^{(*)} \tau \bar\nu$ decays are found to be small after considering other flavour constraints~\cite{Cai:2017wry}.

\section{Observables}
\label{sec:observables}

 In this section, we will introduce the theoretical framework of the relevant flavour processes and discuss the RPV SUSY effects in these processes.

\subsection{$\boldsymbol{b \to c (u) \tau \bar\nu}$ transitions}

With the RPV SUSY contributions, the effective Hamiltonian responsible for $b \to c (u) \tau \bar\nu_\tau$ transitions is given by~\cite{Deshpande:2012rr}
\begin{align}\label{eq:Heff}
\mathcal H_{\rm{eff}} = \frac{4 G_{\rm F}}{\sqrt 2} \sum_{i=u,c} V_{ib} \big( 1 + \mC_{L,i}^\NP \big)\bigl( \bar u_i \gamma ^\mu P_L b  \bigr) \bigl( \bar \tau \gamma _\mu P_L \nu_\tau \bigr),
\end{align}
where tree-level sbottom exchange gives
\begin{align}
\mC_{L,i}^\NP = \frac{v^2}{4m_{\bt_R}^2}\lp3  \sum_{j=1}^3 \lps{j} \left(\frac{V_{ij}}{V_{i3}}\right),
\end{align}
with the Higgs vev $v=246\GeV$. It is noted that this Wilson coefficient is at the matching scale $\mu_{\rm NP} \sim m_{\bt_R}$. However, since the corresponding current is conserved, we can obtain the low-energy Wilson coefficient without considering the Renormalization Group Evolution (RGE) effects, i.e., $\mC_{L,i}^\NP(\mu_b)=\mC_{L,i}^\NP(\mu_\NP)$.

For $b \to c \ell \bar\nu$ transitions, we consider five processes, including $B \to D^{(*)} \ell \bar\nu$~\cite{Celis:2012dk,Bailey:2012jg,Becirevic:2012jf}, $B_c \to \eta_c \ell \bar\nu$~\cite{Wang:2008xt,Murphy:2018sqg}, $B_c \to J/\psi \ell \bar\nu$~\cite{Colangelo:1999zn,Kiselev:2002vz,Hernandez:2006gt,Ivanov:2006ni,Qiao:2012vt,Wen-Fei:2013uea,Hsiao:2016pml,Watanabe:2017mip,Tran:2018kuv}, and $\Lambda_b \to \Lambda _c \ell \bar\nu$~\cite{Gutsche:2015mxa,Shivashankara:2015cta,Dutta:2015ueb,Faustov:2016pal} decays. All these decays, can be uniformly denoted as
\begin{align}\label{eq:decayproc}
M(p_M,\lambda_M)\to N(p_N,\lambda_N)+\ell^{-}(p_\ell,\lambda_\ell)+\bar\nu_\ell(p_{\bar\nu_\ell}),
\end{align}
where $(M,N)=(B, D),\,(B_c,\eta_c)\,,(B, D^*)\,,(B_c, J/\psi)$, and $(\Lambda_b, \Lambda_c)$, and $(\ell ,\bar\nu)=(e,\bar\nu_e),\,(\mu, \bar\nu_\mu)$, and $(\tau, \bar\nu_\tau)$. For each particle $i$ in the above decay, its momentum and helicity are denoted as $p_i$ and $\lambda_i$, respectively. In particular, the helicity of pseudoscalar meson is zero, e.g., $\lambda_D=0$. After summation of the helicity of parent hadron $M$, differential decay width for this process can be written as~\cite{Hagiwara:1989cu,Tanaka:2012nw}
\begin{align}\label{eq:dga}
\text{d}\Gamma^{\lambda_N,\,\lambda_\ell}(M \to N \ell^-\bar\nu_\ell)=\frac{1}{1+2|\lambda_M|}\sum_{\lambda_M} \big\lvert \mathcal{M}^{\lambda_M}_{\lambda_N,\lambda_\ell} \big\rvert^2\frac{\sqrt{Q_+Q_-}}{512\pi^3 m_M^3}\sqrt{1-\frac{m_{\ell}^2}{q^2}}\text{d}q^2 \text{d} \cos\theta_\ell,
\end{align}
where $q= p_M-p_N$, $m_\pm = m_M \pm m_N$, and $Q_\pm =m_\pm^2 - q^2 $. The angle $\theta_\ell\in [0 ,\pi]$ denotes the angle between the three-momentum of $\ell$ and that of $N$ in the $\ell$-$\bar\nu$ center-of-mass frame. With the differential decay width, we can derive the following observables:
\begin{itemize}
\item The decay width and branching ratio
  \begin{align}
    \frac{{\rm d}\mB}{{\rm d}q^2}=\frac{1}{\Gamma_M}\frac{{\rm d}\Gamma}{{\rm d}q^2}=\frac{1}{\Gamma_M}\sum_{\lambda_N,\lambda_\ell}\frac{{\rm d}\Gamma^{\lambda_N,\lambda_\ell}}{{\rm d}q^2},
  \end{align}
  where $\Gamma_{M}$ is the total width of the hadron $M$.
  
\item The LFU ratio
  \begin{align}
  R_{N}(q^2)=\frac{{\rm d}\Gamma(M\to N \tau\bar\nu_\tau)/{\rm d}q^2}{{\rm d}\Gamma(M\to N \ell\bar\nu_\ell)/{\rm d}q^2}\,,\quad
  \end{align}
  where $\text{d}\Gamma(M\to N \ell \bar\nu_\ell)/\text{d}q^2$ in the denominator denotes the average of the different decay widths of the electronic and muonic modes.
  
\item The lepton forward-backward asymmetry
  \begin{align}\label{eq:AFB}
  A_{\rm FB}(q^2) = \frac{\int_{0}^{1} {\rm d}\cos\theta_\ell({\rm d}^2\Gamma/{\rm d}q^2{\rm d}\cos\theta_\ell)-\int_{-1}^{0}{\rm d}\cos\theta_\ell({\rm d}^2\Gamma/{\rm d}q^2{\rm d} \cos\theta_\ell )}{{\rm d}\Gamma/{\rm d}q^2}.
  \end{align}
  
\item The polarization fractions
  \begin{align}\label{eq:PL}
     P_L^{\tau}(q^2)&=\frac{{\rm d}\Gamma^{\lambda_{\tau}=+1/2}/{\rm d}q^2-
                      {\rm d}\Gamma^{\lambda_{\tau}=-1/2}/{\rm d}q^2}{{\rm d}\Gamma/{\rm d}q^2},&&  
    \\
    P_L^N(q^2)&=\frac{{\rm d}\Gamma^{\lambda_N=+1/2}/{\rm d}q^2-
                {\rm d}\Gamma^{\lambda_N=-1/2}/{\rm d}q^2}{{\rm d}\Gamma/{\rm d}q^2},&&(\text{for } N=\Lambda_c)\nonumber
    \\
    P_L^N(q^2)&=\frac{{\rm d}\Gamma^{\lambda_N=0}/{\rm d}q^2}{{\rm d}\Gamma/{\rm d}q^2},&&(\text{for } N=D^*, J/\psi)\nonumber
  \end{align}
\end{itemize}
Explicit expressions of the helicity amplitudes $\mathcal M_{\lambda_N,\lambda_\ell}^{\lambda_M} \equiv \langle N\ell \bar{\nu}_\ell |\mathcal{H}_{\rm eff}|M\rangle$ and all the above observables can be found in ref.~\cite{Sakaki:2013bfa} for $B \to D^{(*)} \tau \bar\nu$ decays, and ref.~\cite{Datta:2017aue} for $\Lambda_b \to \Lambda_c \tau \bar\nu$ decay. The expressions for $B_c \to \eta_c \tau \bar\nu$ and $B_c \to J/\psi \tau \bar\nu$ are analogical to the ones for $B \to D \tau \bar\nu$ and $B \to D^* \tau \bar\nu$, respectively. Since these angular observables are ratios of decay widths, they are largely free of hadronic uncertainties, and thus provide excellent tests of lepton flavour universality. It is noted that the RPV SUSY effects generate operator with the same chirality structure as in the SM, as shown in eq.~(\ref{eq:Heff}). It's straightforward to derive the following relation in all the $b \to c \tau \bar\nu$ decays
\begin{align}\label{eq:b2c:relation}
\frac{R_N}{R_N^\SM} = \left\lvert 1 + \mC_{L,2}^\NP \right\rvert ^2,
\end{align}
for $N=D^{(*)}, \eta_c, J/\psi$, and $\Lambda_c$. Here, vanishing contributions to the electronic and muonic channels are assumed.

The hadronic $M \to N$ transition form factors are important inputs to calculate the observables introduced above. In recent years, notable progresses have been achieved in this field~\cite{Detmold:2015aaa,Gutsche:2015mxa,Bigi:2016mdz,Jaiswal:2017rve,Bernlochner:2017jka,Bigi:2017jbd,Datta:2017aue,Wang:2017jow,Grinstein:2017nlq,Gubernari:2018wyi,Murphy:2018sqg,Berns:2018vpl,Wang:2018duy,Bernlochner:2019ldg,Leljak:2019eyw,FLAG}. For $B \to D^{(*)}$ transitions, it has already been emphasized that the Caprini-Lellouch-Neubert (CLN) parameterization~\cite{Caprini:1997mu} does not account for uncertainties in the values of the subleading Isgur-Wise functions at zero recoil obtained with QCD sum rules~\cite{Neubert:1992wq,Neubert:1992pn,Ligeti:1993hw}, where the number of parameters is minimal~\cite{Bernlochner:2017jka}. In this work, we don’t use such simplified parameterization but adopt the conservative approach in ref.~\cite{Bigi:2016mdz,Bigi:2017jbd}, which is based on the Boyd-Grinstein-Lebed (BGL) parameterization~\cite{Boyd:1997kz}. In addition, we use the $B_c\to \eta_c,\,J/\psi$ transition form factors obtained in the covariant light-front approach~\cite{Wang:2008xt}. For the $\Lambda_b \to \Lambda_c$ transition form factor, we adopt the recent Lattice QCD results in ref.~\cite{Detmold:2015aaa,Datta:2017aue}. Explicit expressions of all the form factors used in our work are recapitulated in appendix~\ref{sec:form factor}.

For $b \to u \tau \bar\nu$ transitions, we consider $B \to \tau \bar\nu$, $B \to \pi \tau\bar\nu$ and $B \to \rho \tau \bar\nu$ decays. Similar to eq.~(\ref{eq:b2c:relation}), we have
\begin{align}
  \frac{\mB(B \to \tau \bar\nu)}{\mB(B \to \tau \bar\nu)_\SM}=
  \frac{\mB(B \to \pi \tau \bar\nu)}{\mB(B \to \pi \tau \bar\nu)_\SM}=
  \frac{\mB(B \to \rho \tau \bar\nu)}{\mB(B \to \rho \tau \bar\nu)_\SM}=
  \left\lvert 1 + \mC_{L,1}^\NP \right\rvert ^2.
\end{align}
It is noted that the SUSY contributions to both $ b \to u \tau \bar\nu $ and $b \to c \tau \bar\nu$ transitions depend on the same set of parameters, $\lp1$, $\lp2$, and $\lp3$. Therefore, the ratios $R_{D^{*}}$ are  related to the $B \to \tau \bar\nu$ decay.

\subsection{Other processes}

The Flavour-Changing Neutral Current (FCNC) decays $B^+ \to K^+ \nu \bar\nu$ and $B^+ \to \pi^+ \nu \bar\nu$ are induced by the $b \to s \nu \bar\nu$ and $b \to d \nu \bar\nu$ transitions, respectively. In the SM, they are forbidden at the tree level and highly suppressed at the one-loop level due to the GIM mechanism. In the RPV SUSY, the sbottoms can contribute to these decays at the tree level, which result in strong constraints on the RPV couplings. Similar to the $b \to c (u) \tau \bar\nu$ transitions, the RPV interactions do not generate new operators beyond the ones presented in the SM. Therefore, we have~\cite{Altmannshofer:2017poe,Deshpand:2016cpw}
\begin{align}
 \frac{\mB(B^+ \to K^+ \nu \bar\nu)}{\mB(B^+ \to K^+ \nu \bar\nu)_\text{SM}} &= \frac{2}{3} + \frac{1}{3} \bigg\lvert 1 - \frac{v^2}{2m^2_{\bt_R}} \frac{\pi s_W^2}{\alpha_\text{em}} \frac{\lp3\lps2 }{V_{tb} V_{ts}^*} \frac{1}{X_t}  \bigg\rvert^2,
  \\
 \frac{\mB(B^+ \to \pi^+ \nu \bar\nu)}{\mB(B^+ \to \pi^+ \nu \bar\nu)_\text{SM}} &= \frac{2}{3} + \frac{1}{3} \bigg\lvert 1 - \frac{v^2}{2m^2_{\bt_R}} \frac{\pi s_W^2}{\alpha_\text{em}} \frac{\lp3 \lps1}{V_{tb} V_{td}^*} \frac{1}{X_t}  \bigg\rvert^2, \nonumber
\end{align}
where the gauge-invariant function $X_t = 1.469\pm0.017$ arises from the box and $Z$-penguin diagrams in the SM~\cite{Buras:2014fpa}.

The leptonic $W$ and $Z$ couplings are also important to probe the RPV SUSY effects~\cite{Feruglio:2016gvd,Feruglio:2017rjo}. In particular, $W$ and $Z$ couplings involving left-handed $\tau$ leptons can receive contributions from the loop diagrams mediated by top quark and sbottom. These effects modify the leptonic $W$ and $Z$ couplings as~\cite{Altmannshofer:2017poe}
\begin{align}\label{eq:Z coupling}
 \frac{g_{Z\tau_L\tau_L}}{g_{Z\ell_L\ell_L}} =& 1 - \frac{3 |\lp3|^2}{16\pi^2} \frac{1}{1 - 2 s_W^2} \frac{m_t^2}{m^2_{\bt_R}} f_Z\bigg(\frac{m_t^2}{m^2_{\bt_R}}\bigg) , \\
 \frac{g_{W\tau_L\nu_\tau}}{g_{W\ell_L\nu_\ell}} =& 1 - \frac{3 |\lp3|^2}{16\pi^2} \frac{1}{4} \frac{m_t^2}{m^2_{\tilde b_R}} f_W\bigg(\frac{m_t^2}{m^2_{\bt_R}}\bigg) , \nonumber
\end{align}
where $\ell = e, \mu$ and $s_W=\sin\theta_W$ with $\theta_W$ the weak mixing angle. The loop functions $f_Z(x)$ and $f_W(x)$ have been calculated in refs.~\cite{Feruglio:2016gvd,Feruglio:2017rjo,Altmannshofer:2017poe} and are given by $f_Z(x) = 1/(x-1) - \log x/(x-1)^2$ and $f_W(x) = 1/(x-1) - (2-x)\log x/(x-1)^2$. Experimental measurements on the $Z \tau_L \tau_L$ couplings have been performed at the LEP and SLD~\cite{ALEPH:2005ab}. Their combined results give $g_{Z\tau_L\tau_L} / g_{Z\ell_L\ell_L} = 1.0013 \pm 0.0019 $~\cite{Altmannshofer:2017poe}. The $W \tau_L \nu_\tau$ coupling can be extracted from  $\tau$ decay data. The measured $\tau$ decay fractions compared to the $\mu$ decay fractions give $ g_{W\tau_L\nu_\tau} / g_{W\ell_L\nu_\ell} = 1.0007 \pm 0.0013$~\cite{Altmannshofer:2017poe}. Both the leptonic $W$ and $Z$ couplings are measured at few permille level. Therefore, they will put strong bounds on the RPV coupling $\lp3$.

The RPV interactions can also affect $K$-meson decays, e.g., $K \to \pi \nu \bar\nu$, $D$-meson decays, e.g., $D\to \tau \bar\nu$, and $\tau$ lepton decays, e.g., $\tau \to \pi \nu$. However, as discussed in ref.~\cite{Altmannshofer:2017poe}, their constraints are weaker than the ones from the processes discussed above. In addition, bound from the $B_c$ lifetime~\cite{Li:2016vvp,Alonso:2016oyd} is not relevant, since the RPV SUSY contributions to $B_c \to \tau \bar\nu$ are not chirally enhanced compared to the SM.

Another interesting anomalies arise in the recent LHCb measurements of $R_{K^{(*)}}\equiv \mB(B \to K^{(*)} \mu^+ \mu^-)/\mB(B \to K^{(*)} e^+ e^-)$, which show about $2\sigma$ deviation from the SM prediction~\cite{Aaij:2014ora,Aaij:2017vbb} and are refered to as $R_{K^{(*)}}$ anomalies. The $R_{K^{(*)}}$ anomalies imply hints of LFU violation in $b \to s \ell^+ \ell^-$ transition. In the RPV SUSY, the left-handed stop can affect this process at the tree level, and the right-handed sbottom can contribute at the one-loop level. However, as discussed in ref.~\cite{Deshpand:2016cpw}, once all the other flavour constraints are taken into account, no parameter space in the RPV SUSY can explain the current $R_{K^{(*)}}$ anomaly.

Finally, we briefly comment the direct searches for the sbottoms at the LHC. Using data corresponding to $35.9\,\text{fb}^{-1}$ at 13\,TeV, the CMS collaboration has performed search for heavy scalar leptoquarks in $p p \to t \bar t \tau^+ \tau^-$ channel. The results can be directly reinterpreted in the context of pair produced sbottoms decaying into top quark and $\tau$ lepton pairs via the RPV coupling $\lp3$. Then, the mass of the sbottom is excluded up to $810\GeV$ at 95\%\,CL~\cite{Sirunyan:2018nkj}.

\section{Numerical results and discussions}
\label{sec:Numerical results}

\begin{table}[t]
  \centering
  \setlength{\extrarowheight}{5pt}
  \begin{tabular}{cccc}
    \toprule
    Input & Value & Unit & Ref
    \\
    \midrule
    $m_t^{\rm pole}$ & $173.1 \pm 0.9$ & GeV & \cite{Tanabashi:2018oca}
    \\
    $m_b(m_b)$ & $4.18\pm0.03$ & GeV &  \cite{Tanabashi:2018oca}
    \\
    $m_c(m_c)$ & $1.28\pm0.03$ & GeV &\cite{Tanabashi:2018oca}
    \\
    $A$ & $0.8396_{-0.0298}^{+0.0080}$ & &\cite{CKMfitter}
    \\
    $\lambda$ & $ 0.224756^{+0.000163}_{-0.000065}$ & & \cite{CKMfitter}
    \\
    $\bar\rho$ & $ 0.123^{+0.023}_{-0.023} $ & & \cite{CKMfitter}
    \\
    $\bar\eta$ & $ 0.375^{+0.022}_{-0.017}$ & & \cite{CKMfitter}
    \\
    \bottomrule
  \end{tabular}
  \caption{Input parameters used in our numerical analysis.}
  \label{tab:inputs}
\end{table}

In this section, we proceed to present our numerical analysis for the RPV SUSY scenario introduced in section~\ref{sec:SUSY}. We will derive constraints on the RPV couplings and study their effects to various processes.

The most relevant input parameters used in our numerical analysis are presented in table~\ref{tab:inputs}. Using the theoretical framework described in section~\ref{sec:observables}, the SM predictions for the $B \to D^{(*)} \tau \bar\nu$, $B_c \to \eta_c \tau \bar\nu$, $B_c \to J/\psi \tau \bar\nu$, and $\Lambda_b \to \Lambda_c \tau \bar\nu$ decays are given in table \ref{tab:numerics}. In order to obtain the theoretical uncertainties, we vary each input parameter within its $1\sigma$ range and add each individual uncertainty in quadrature. For the uncertainties induced by form factors, we will also include the correlations among the fit parameters. In particular, for the $\Lambda_b \to \Lambda_c \tau \bar\nu$ decay, we follow the treatment of ref.~\cite{Detmold:2015aaa} to obtain the statistical and systematic uncertainties induced by the form factors. From table~\ref{tab:numerics}, we can see that the experimental data on the ratios $R_D$, $R_{D^*}$ and $R_{J/\psi}$ deviate from the SM predictions by $2.33\sigma$, $2.74\sigma$ and $1.87\sigma$, respectively.

\begin{table}[t]
  \centering
  \setlength{\extrarowheight}{5pt}
  \begin{tabular}{c c c c c}
        \toprule
    Observable & Unit & SM  & RPV SUSY & Exp.
    \\
    \midrule
    $\mB(B \to \tau \bar\nu)$ & $10^{-4}$ &$0.947 ^{+0.182}_{-0.182} $ & $[0.760,1.546]$  &  $ 1.44 \pm 0.31 $~\cite{HFLAV}
    \\
    $\mB (B^+ \to \pi ^+ \nu \bar\nu)$ & $10^{-6}$ & $0.146 ^{+0.014}_{-0.014}$ & $[0.091 ,14.00]$ & $< 14 $~\cite{Tanabashi:2018oca}
    \\
    $\mB (B^+ \to K^+ \nu \bar\nu)$ & $10^{-6}$ &$3.980 ^{+0.470}_{-0.470}$ & $[6.900,16.00] $ & $<16$~\cite{Tanabashi:2018oca}
    \\
    \midrule
    $\mB(B\to D\tau\bar{\nu})$ & $10^{-2}$ & $0.761 ^{+0.021}_{-0.055}$ & $[0.741,0.847]$  & $0.90 \pm 0.24$ \cite{Tanabashi:2018oca}
    \\
    $R_{D}$ & & $0.300^{+0.003}_{-0.003}$  & $[0.314,0.330]$ & $0.407\pm0.039\pm0.024 $~\cite{HFLAV}
    \\
    $\mB(B_{c}\to \eta_{c}\tau\bar{\nu})$ &  $10^{-2}$  &  $0.219 ^{+0.023}_{-0.029}$   & $ [0.199, 0.262] $  & \rule{2em}{1pt}
    \\
    $R_{\eta_{c}}$ & & $0.280^{+0.036}_{-0.031}$  & $[0.262,0.342]$ & \rule{2em}{1pt}
    \\
    $\mB(B\to D^{*}\tau\bar{\nu})$&  $10^{-2}$  &  $1.331^{+0.103}_{-0.122}$   & $[1.270,1.554]$  & $1.78 \pm 0.16 $~\cite{Tanabashi:2018oca}
    \\
    $R_{D^{*}}$ & & $0.260^{+0.008}_{-0.008}$ & $ [0.267,0.291]$ & $0.306\pm0.013\pm0.007 $~\cite{HFLAV}
    \\
    $P_L^\tau$ & & $-0.467^{+0.067}_{-0.061}$\hphantom{$-$}  &$ [-0.528,-0.400]$ &$ -0.38\pm0.51^{+0.21}_{-0.16} $~\cite{Hirose:2016wfn,Hirose:2017dxl}
    \\
    $P_L^{D^*}$ & & $0.413^{+0.032}_{-0.031}$ & $ [0.382,0.445]$&$0.60 \pm 0.08 \pm 0.04 $~\cite{Adamczyk:2019wyt,Abdesselam:2019wbt}
    \\
    $\mB(B_{c}\to J/\psi\tau\bar{\nu})$ &  $10^{-2}$  & $0.426^{+0.046}_{-0.058}$ & $ [0.387, 0.512]$  & \rule{2em}{1pt}
    \\
    $R_{J/\psi}$ & & $0.248^{+0.006}_{-0.006}$ &  $[0.254,0.275]$  & $0.71\pm0.17\pm0.18 $~\cite{Aaij:2017tyk}
    \\
    $\mB(\Lambda_{b}\to\Lambda_{c}\tau\bar{\nu})$ &  $10^{-2}$  &  $1.886 ^{+0.107}_{-0.165}$ & $[1.807,2.159]$  & \rule{2em}{1pt}
    \\
    $R_{\Lambda_{c}}$ & & $0.332^{+0.011}_{-0.011}$ &  $[0.337,0.372]$ & \rule{2em}{1pt}
    \\
    \bottomrule
  \end{tabular}
  \caption{Predictions for the branching fractions and the ratios $R$ of the five $b \to c \tau \bar\nu$ channels in the SM and RPV SUSY. The sign ``\rule{2em}{1pt}'' denotes no available measurements at present. Upper limits are all at 90\% CL.}
  \label{tab:numerics}
\end{table}

\subsection{Constraints}

In the RPV SUSY scenario introduced in section~\ref{sec:SUSY}, the relevant parameters to explain the $R_{D^{(*)}}$ anomalies are $(\lp1 ,\, \lp2 ,\, \lp3)$ and $m_{\bt_R}$. In section~\ref{sec:observables}, we know only the three products of the RPV couplings, $(\lp1\lps3 ,\, \lp2\lps3,\, \lp3\lps3 )$, appear in the various flavour processes. In the following analysis, we will assume these products are real and derive bounds on them. We impose the experimental constraints in the same way as in refs.~\cite{Jung:2012vu,Chiang:2017etj}; i.e., for each point in the parameter space, if the difference between the corresponding theoretical prediction and experimental data is less than $2\sigma$ $(3\sigma)$ error bar, which is evaluated by adding the theoretical and experimental errors in quadrature, this point is regarded as allowed at $2\sigma$ $(3\sigma)$ level. From section~\ref{sec:observables}, it is known that the RPV couplings always appear in the form of $\lambda_{3i3}^\prime \lps3/m_{\bt_R}^2$ in all the $B$ decays. Therefore, we can take $m_{\bt_R}=1\TeV$ without loss of generality, which is equivalent to absorb $m_{\bt_R}$ into $\lp{i}\lps3$. Furthermore, the choice of $m_{\bt_R}=1\TeV$ is compatible with the direct searches for the sbottoms at CMS~\cite{Sirunyan:2018nkj}. In the SUSY contributions to the couplings $g_{Z\tau_L\tau_L}$ and $g_{W\tau_L\nu_\tau}$ in eq.~(\ref{eq:Z coupling}), additional $m_{\bt_R}$ dependence arises in the loop functions $f_Z(m_t^2/m_{\bt_R}^2)$ and $f_W(m_t^2/m_{\bt_R}^2)$, respectively. As can be seen in the next subsection, our numerical results show such $m_{\bt_R}$ dependence is weak and the choice of $m_{\bt_R}=1\TeV$ does not lose much generality.

\begin{figure}[t]
  \captionsetup[subfigure]{labelformat=empty}
  \centering
  \begin{subfigure}[b]{0.4\textwidth}
    \includegraphics[width=\textwidth]{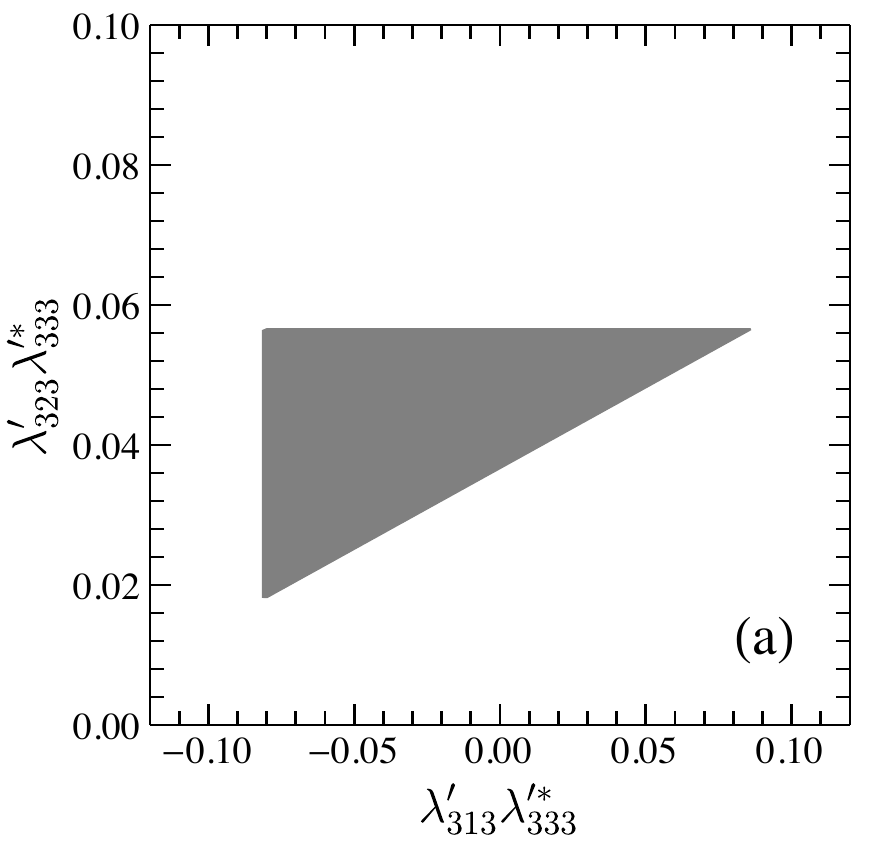}
    \caption{}
    \label{fig:lp1:lp2}
  \end{subfigure}
  \qquad
  \begin{subfigure}[b]{0.4\textwidth}
    \includegraphics[width=\textwidth]{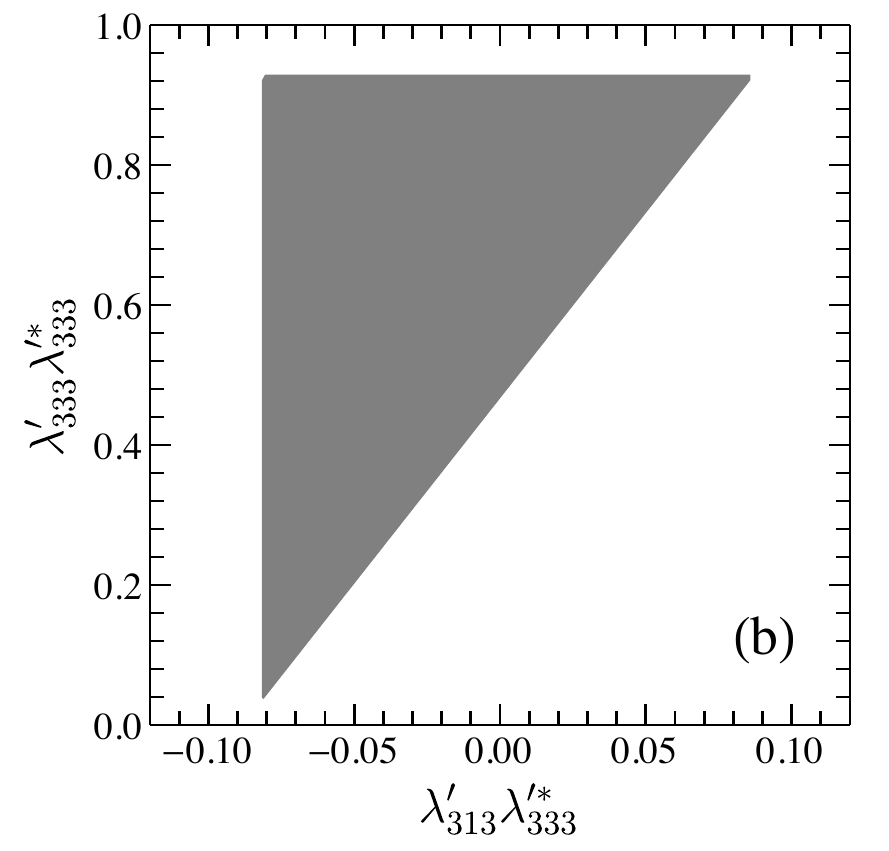}
    \caption{}
    \label{fig:lp1:lp3}
  \end{subfigure}
  \\\vspace{-1em}
  \begin{subfigure}[b]{0.4\textwidth}
    \includegraphics[width=\textwidth]{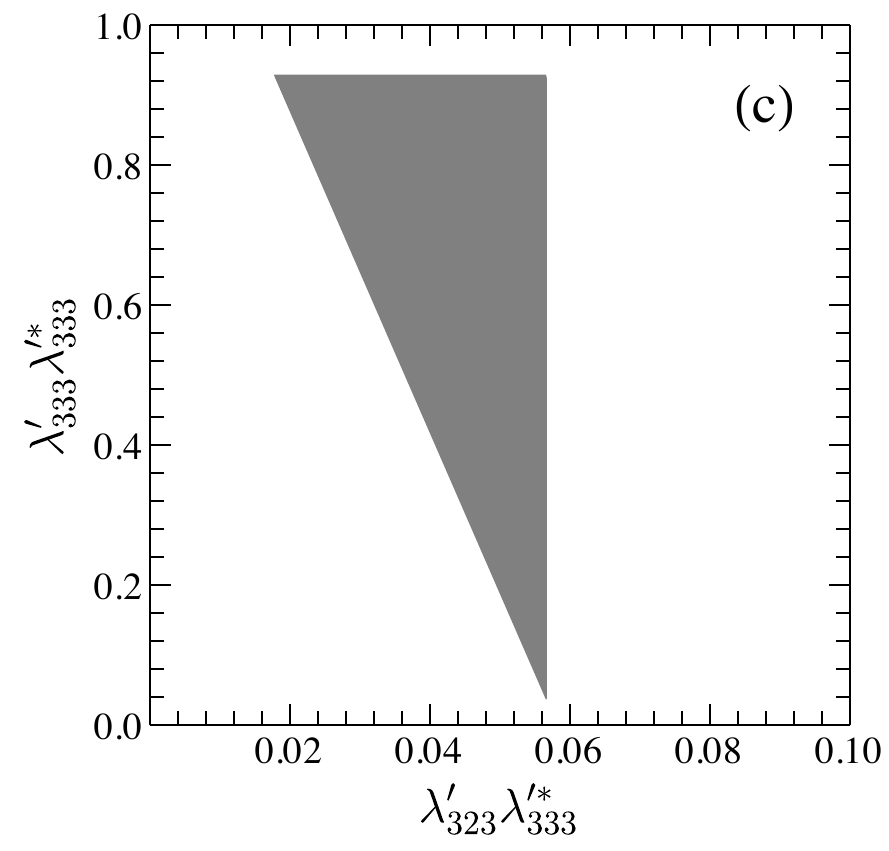}
    \caption{}
    \label{fig:lp2:lp3}
  \end{subfigure}
  \qquad
  \begin{subfigure}[b]{0.4\textwidth}
    \includegraphics[width=\textwidth]{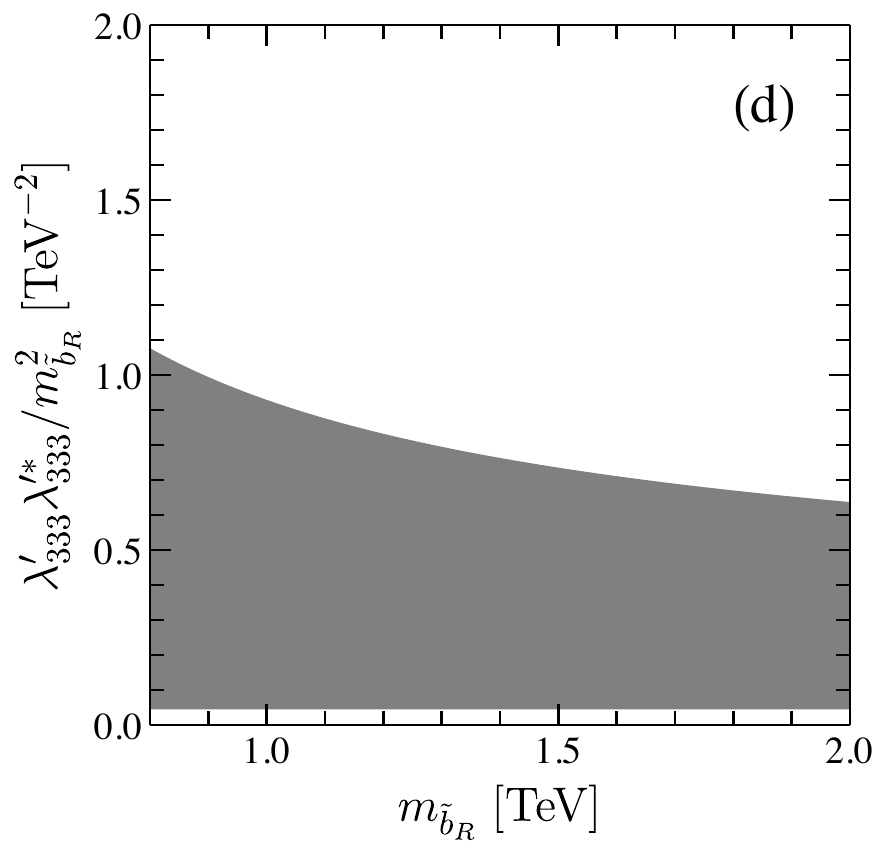}
    \caption{}
    \label{fig:mb:lp3}
  \end{subfigure}
  \vspace{-1.5em}
  \caption{Allowed parameter space of $(\lp1\lps3 ,\, \lp2\lps3 ,\, \lp3\lps3)$ by all the flavour processes at $2\sigma$ level with $m_{\bt_R}=1\TeV$, plotted in the $(\lp1\lps3 ,\, \lp2\lps3)$ (\textbf{a}), $(\lp1\lps3 ,\, \lp3\lps3)$ (\textbf{b}), and $(\lp2\lps3,\,\lp3\lps3)$ (\textbf{c}) plane. Figure \textbf{d} shows the allowed region in $(m_{\bt_R}, \lp3\lps3/m_{\bt_R}^2)$ plane.}
  \label{fig:parameter space}
\end{figure}

As shown in table~\ref{tab:numerics}, the current experimental upper bounds on the branching ratio of $B^+ \to K^+ \nu \bar\nu$ and $B^+ \to \pi ^+ \nu \bar\nu$ are one order above their SM values. However, since the SUSY contributes to these decays at the tree level, the RPV couplings are strongly constrained as
\begin{align}\label{eq:l1l2}
  -0.082 <& \lp1\lps3 < 0.090, \quad(\text{from} \quad B^+ \to \pi^+ \nu \bar\nu)
  \\
  -0.098 <& \lp2\lps3 < 0.057, \quad(\text{from} \quad B^+ \to K^+ \nu\bar\nu) \nonumber
\end{align}
at $2\sigma$ level. For the leptonic $W$ and $Z$ couplings, the current measurements on $g_{W\tau_L \nu_\tau}/ g_{W \ell_L \nu_\ell}$ and $g_{Z\tau_L \tau_L}/ g_{Z \ell_L \ell_L}$ have achieved to the precision of few permille. We find that the latter can give stronger constraint, which reads
\begin{align}\label{eq:l3}
   \hphantom{-0.000<} & \lp3\lps3 < 0.93 , \quad(\text{from} \quad g_{Z \tau_L\tau_L}/g_{Z \ell_L \ell_L})
\end{align}
or $ |\lp3| < 0.96$, at $2\sigma$ level. It is noted that this upper bound prevents the coupling $\lp3$ from developing a Landau pole below the GUT scale~\cite{Barger:1995qe}.

As discussed in section~\ref{sec:observables}, the RPV interactions affect $b \to c \tau \bar\nu$ transitions via the three products $(\lp1\lps3 ,\, \lp2\lps3 ,\, \lp3\lps3)$. After considering the above individual constraints at $2\sigma$ level, parameter space to explain the current measurements on $R_{D^{(*)}}$, $R_{J/\psi}$, $P_L^\tau (D^*)$ and $P_L^{D^*}$ are shown in figure~\ref{fig:parameter space} for $m_{\bt_R}=1\TeV$. We can see that the $B \to D^{(*)} \tau \bar\nu$ decays and other flavour observables put very stringent constraints on the RPV couplings. The combined constraints are slightly stronger than the individual ones in eq.~(\ref{eq:l1l2}) and (\ref{eq:l3}). It is also noted that, after taking into account the bounds from $B^+ \to K^+ \nu \bar\nu$ and $g_{Z\tau_L\tau_L}$, the $ B \to D^{(*)} \tau \bar\nu$ decays are very sensitive to the product $\lp2\lps3$. As a result, the current $R_{D^{(*)}}$ anomalies give a lower bound on $|\lp2\lps3|$. Finally, the combined bounds in figure~\ref{fig:parameter space} read numerically,
\begin{align}
  -0.082 &< \lp1 \lps3 < 0.087, \quad(\text{from combined constraints})
  \\
  0.018 & < \lp2 \lps3 < 0.057, & & \nonumber
  \\
  0.033 & < \lp3 \lps3 < 0.928. & & \nonumber
\end{align}
As can be seen, a weak lower bound on $\lp3\lps3$ is also obtained. In addition, although the constraints from the $D^*$ polarization fraction $P_L^{D^*}$ are much stronger than the ones from the $\tau$ polarization fraction $P_L^\tau$, this observable can't provide further constraints on the RPV couplings. Due to the previous discussions, we show the combined upper bound on $\lp3\lps3/m_{\bt_R}^2$ as a function of $m_{\bt_R}$ in figure~\ref{fig:mb:lp3}. It can be seen that the upper limit of $\lp3\lps3/m_{\bt_R}^2$ just changes around 20\% by varying $m_{\bt_R}$ from $800\GeV$ to $2000\GeV$. Therefore, one can approximately obtain the allowed parameter space for $m_{\bt_R}\neq 1\TeV$ from figure~\ref{fig:lp1:lp2}-\ref{fig:lp2:lp3} by timing a factor of $(m_{\bt_R}/1\TeV)^2$.

\subsection{Predictions}

\begin{figure}[t]
  \captionsetup[subfigure]{labelformat=empty}
  \centering
  \begin{subfigure}[b]{0.3\textwidth}
    \includegraphics[width=\textwidth]{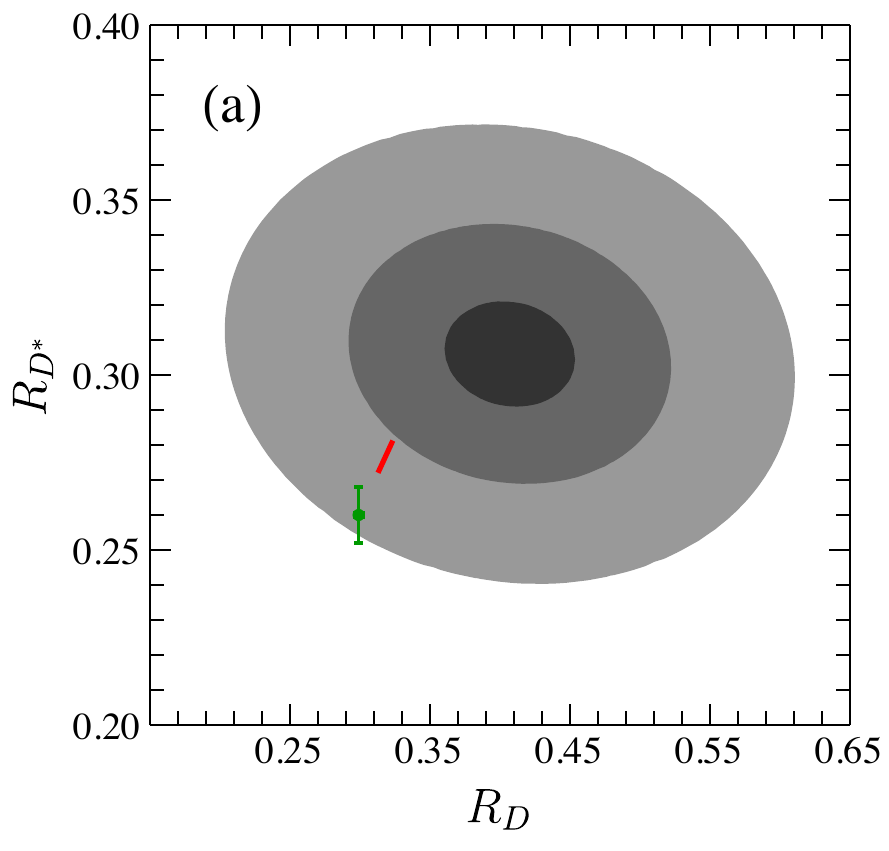}
    \caption{}
    \label{fig:correlation:1}
  \end{subfigure}
  \quad
  \begin{subfigure}[b]{0.3\textwidth}
    \includegraphics[width=\textwidth]{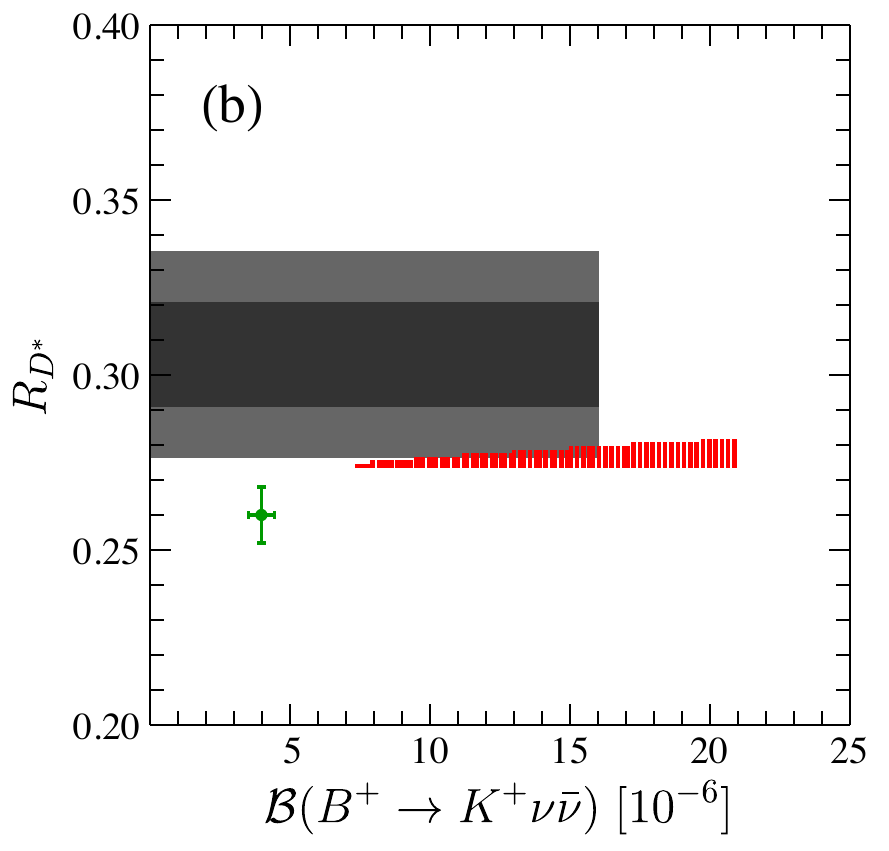}
    \caption{}
    \label{fig:correlation:2}
  \end{subfigure}
  \quad
  \begin{subfigure}[b]{0.3\textwidth}
    \includegraphics[width=\textwidth]{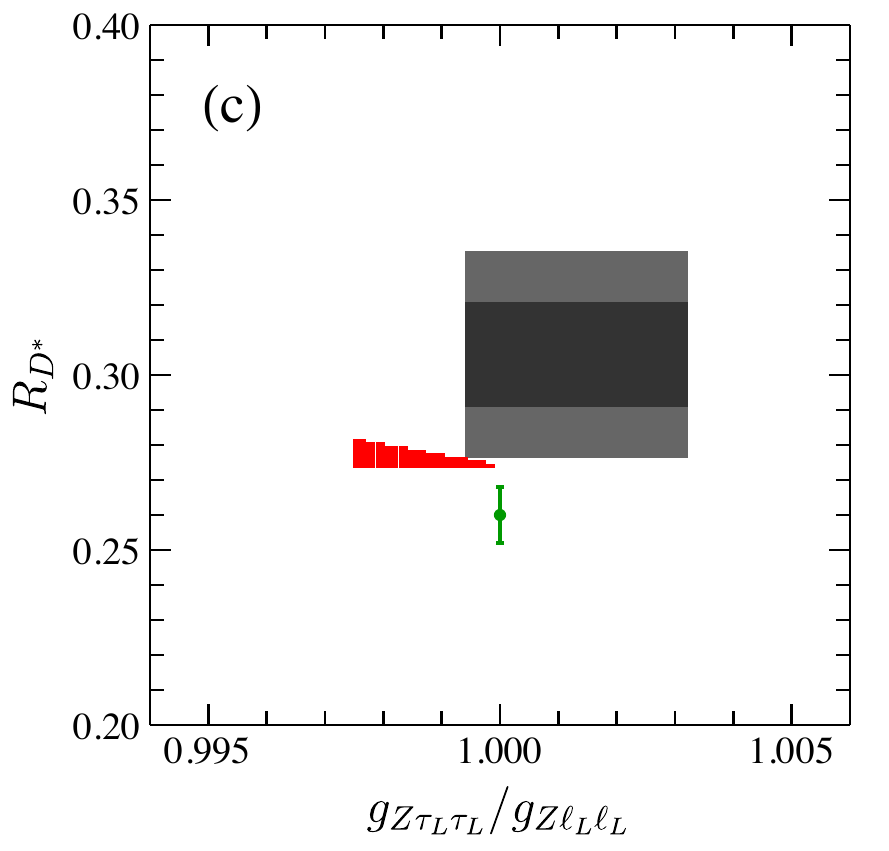}
    \caption{}
    \label{fig:correlation:3}
  \end{subfigure}
  \\[-1em]
  \begin{subfigure}[b]{0.3\textwidth}
    \includegraphics[width=\textwidth]{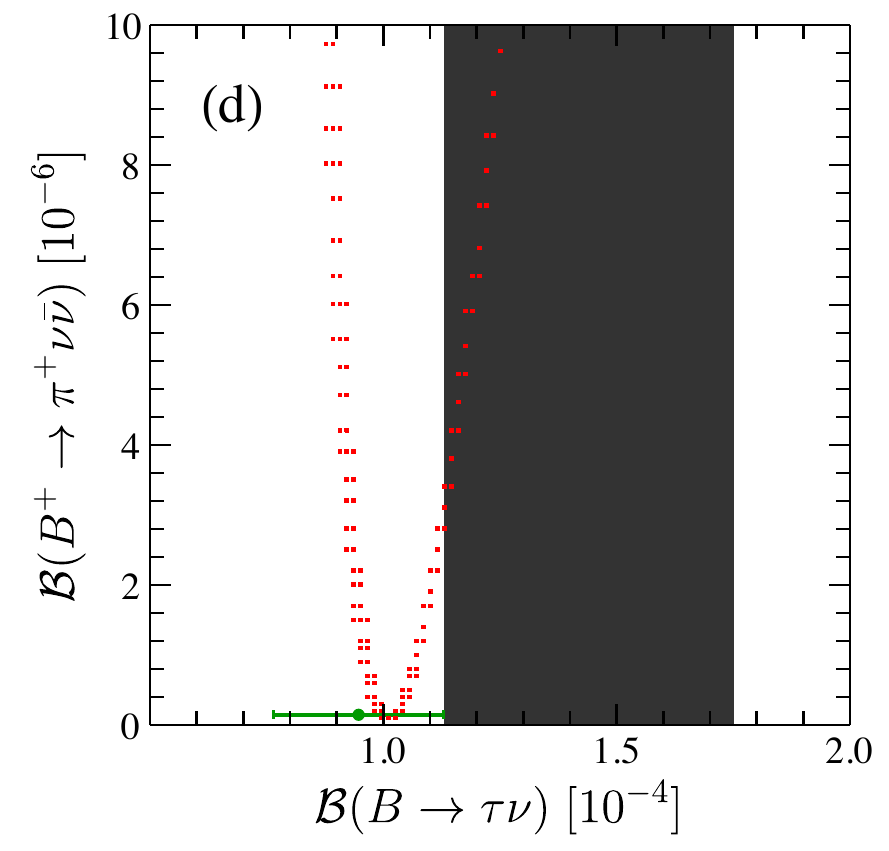}
    \caption{}
    \label{fig:correlation:4}
  \end{subfigure}
  \quad
  \begin{subfigure}[b]{0.3\textwidth}
    \includegraphics[width=\textwidth]{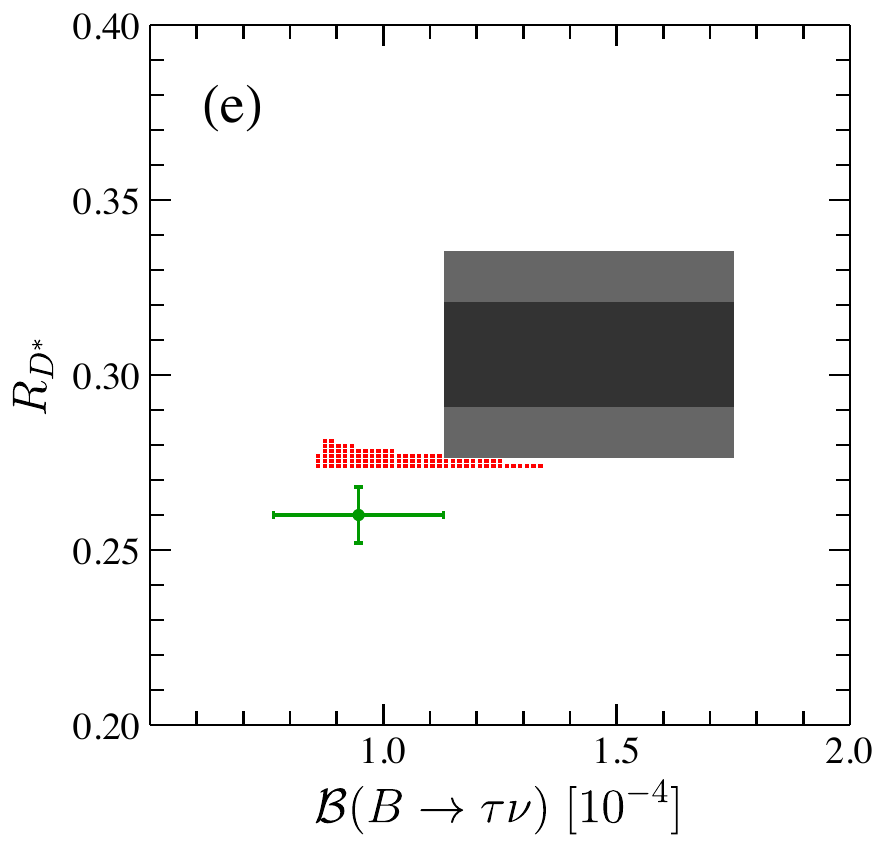}
    \caption{}
    \label{fig:correlation:5}
  \end{subfigure}
  \quad
  \begin{subfigure}[b]{0.3\textwidth}
    \includegraphics[width=\textwidth]{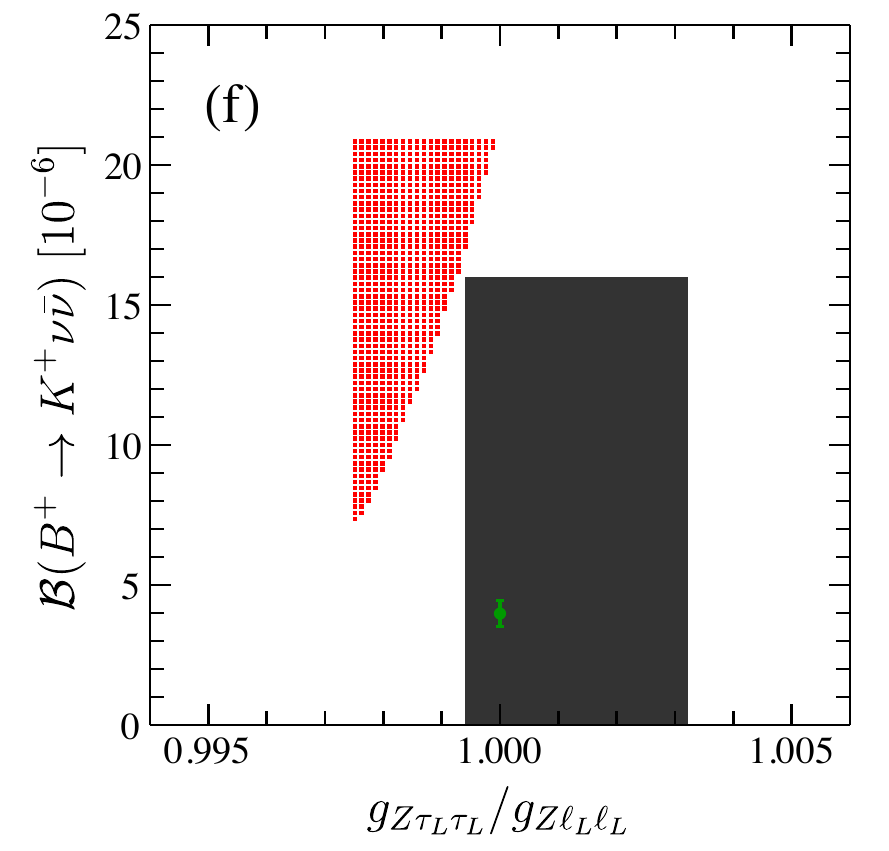}
    \caption{}
    \label{fig:correlation:6}
  \end{subfigure}
  \vspace{-1.5em}
  \caption{Correlations among various observables. The SM predictions correspond to the green cross, while the correlations in the RPV SUSY are shown in red points. In figure \ref{fig:correlation:1}, the current HFLAV averages for $R_D$ and $R_{D^*}$ are shown in black region, and the $2\sigma$ ($4\sigma$) experimental region is shown in gray (light gray) region. In the other figures, the $1\sigma$ experimental region is shown in black. The $2 \sigma$ regions for $R_{D^*}$ is also given in gray.}
  \label{fig:correlation}
\end{figure}

In the parameter space allowed by all the constraints at $2\sigma$ level, correlations among several observables are obtained, which are shown in figure~\ref{fig:correlation}. In these figures, the SUSY predictions are central values without theoretical uncertainties. From figure~\ref{fig:correlation:1}, we can see that the central value of $R_D$ and $R_{D^*}$ are strongly correlated, as expected in eq.~(\ref{eq:b2c:relation}). It is noted that the SUSY effects can only enhance the central value of $R_{D^{(*)}}$ by about 8\%, so that $R_{D^{(*)}}$ approach to, but still lie outside, the $2\sigma$ range of the HFLAV averages. Therefore, future refined measurements will provide a crucial test on the RPV SUSY explaination of $R_{D^{(*)}}$ anomalies. At Belle II, precisions of $R_{D^{(*)}}$ measurements are expected to be about 2-4\%~\cite{Kou:2018nap} with a luminosity of $50\ab^{-1}$. From figure~\ref{fig:correlation:2}, it can be seen that both $R_{D^*}$ and $\mB (B^+ \to K^+ \nu \bar\nu)$ deviate from their SM predictions. The lower bound for the latter is $\mB (B^+ \to K^+ \nu \bar\nu)> 7.37 \times 10^{-6}$, which is due to the lower bound on $\lp2 \lps3 > 0.018$ obtained in the last section. Compared to the SM prediction $\mB (B^+ \to K^+ \nu \bar\nu)_\SM = (3.98 \pm 0.47) \times 10^{-6}$, such significant enhancement makes this decay an important probe of the RPV SUSY effects. In the future, Belle II with $50 \ab^{-1}$ data can measure its branching ratio with a precision of 11\%~\cite{Kou:2018nap}. Another interesting correlation arises between $\mB (B ^+ \to K^+ \nu \bar\nu )$ and $g_{Z\tau_L\tau_L}/g_{Z\ell_L\ell_L}$. As shown in figure~\ref{fig:correlation:6}, the RPV SUSY effects always enhance $\mB(B^+ \to K^+ \nu \bar\nu) $ and suppress $g_{Z\tau_L\tau_L}/g_{Z\ell_L\ell_L}$ simultaneously. When $g_{Z\tau_L\tau_L}/g_{Z\ell_L\ell_L}$ approaches to the SM value $1$, the branching ratio of $B^+ \to K^+ \nu \bar\nu$ maximally deviates from its SM prediction. In figure~\ref{fig:correlation:4} and \ref{fig:correlation:5}, we show the correlations involving $B \to \tau \nu$ decay. It can be seen that the SUSY prediction on $\mB( B \to \tau \bar\nu)$ almost lies in the SM $1\sigma$ range. Since the future Belle II sensitivity at $50\ab^{-1}$ is comparable to the current theoretical uncertainties~\cite{Kou:2018nap}, much more precise theoretical predictions are required in the future to probe the SUSY effects. 

\begin{figure}[t]
  \centering
  \includegraphics[width=0.4\textwidth]{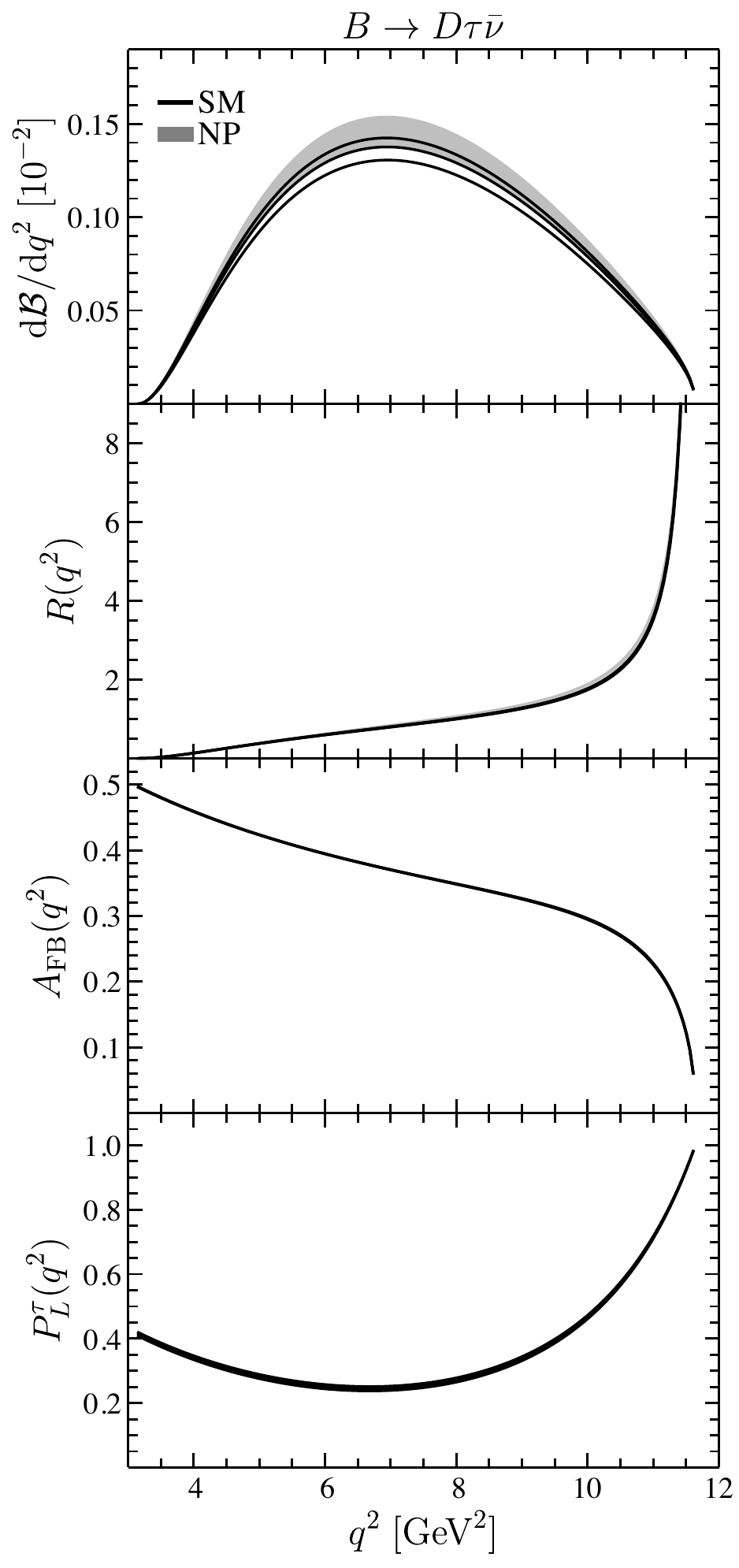}
  \qquad
  \includegraphics[width=0.4\textwidth]{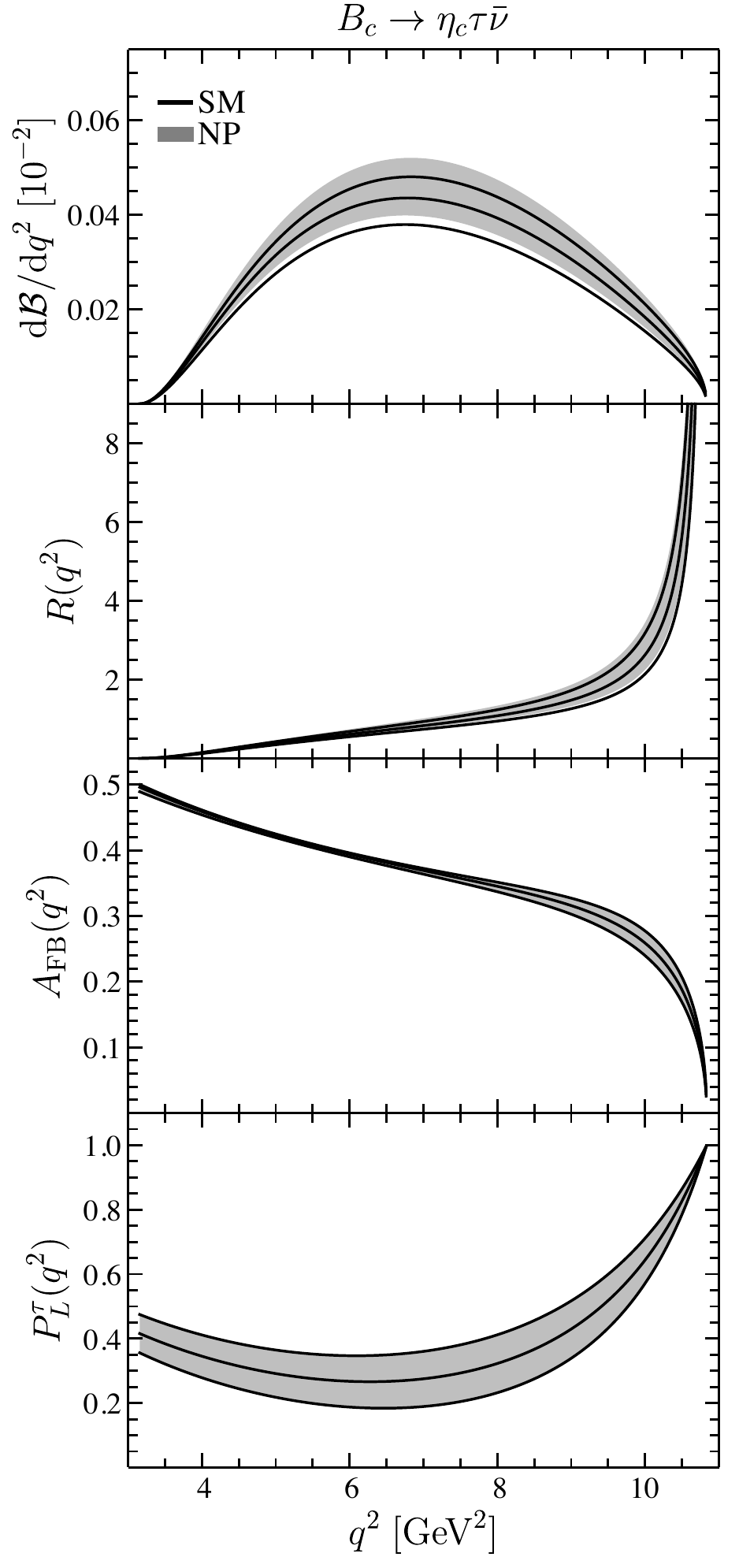}  
  \caption{Differential observables in $B \to D \tau \bar\nu$ (left) and $B_c \to \eta_c \tau \bar\nu$ (right) decays. The black curves (gray band) indicate the SM (SUSY) central values with $1\sigma$ theoretical uncertainty.}
  \label{fig:B}
\end{figure}

\begin{figure}[t]
  \centering
  \includegraphics[width=0.4\textwidth]{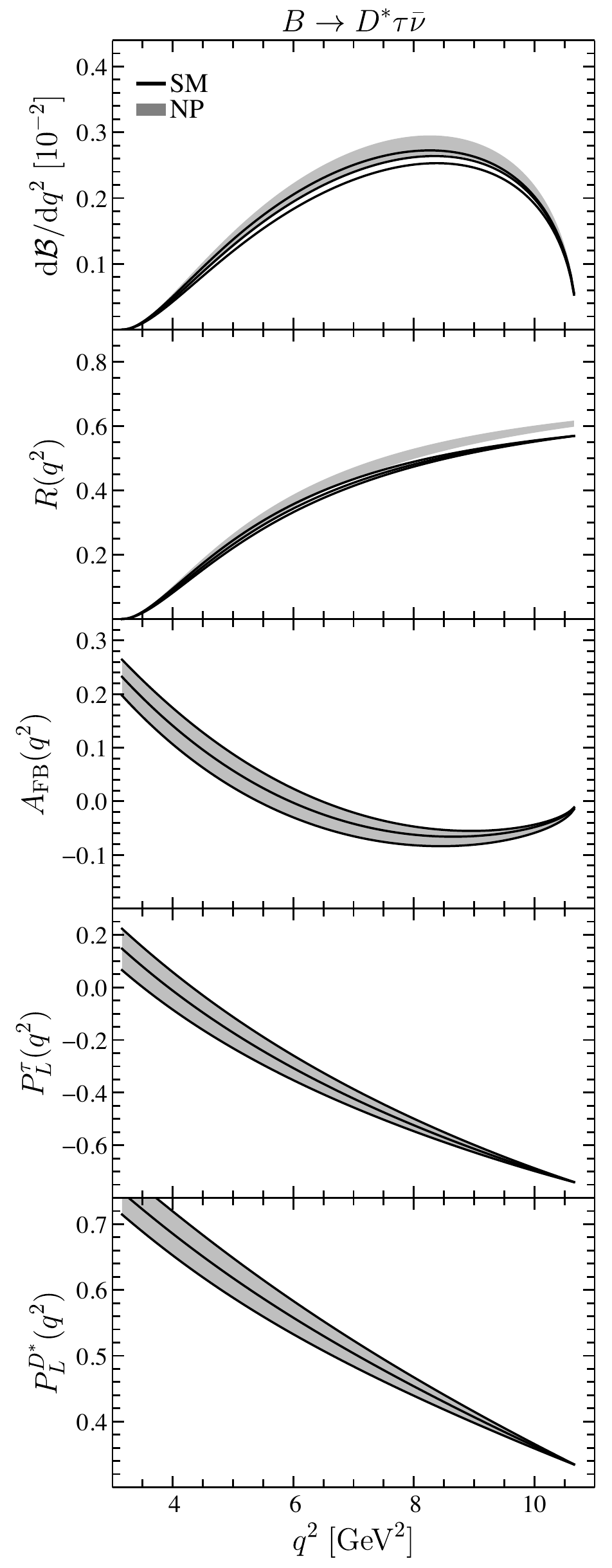}
  \qquad
  \includegraphics[width=0.4\textwidth]{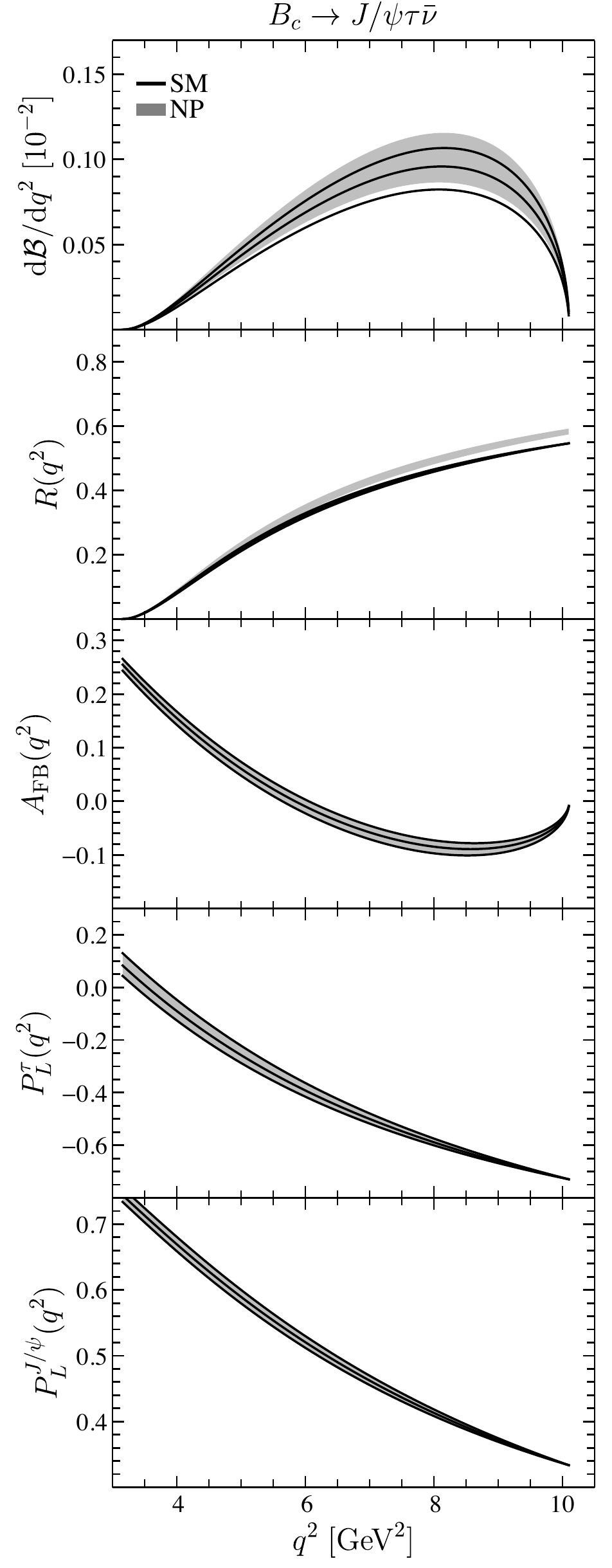}
  \caption{Differential observables in $B \to D^* \tau \bar\nu$ (left) and $B_c \to J/\psi \tau\bar\nu$ (right) decays. Other captions are the same as in figure~\ref{fig:B}.}
  \label{fig:Bc}
\end{figure}

\begin{figure}[t]
  \centering
  \includegraphics[width=0.4\textwidth]{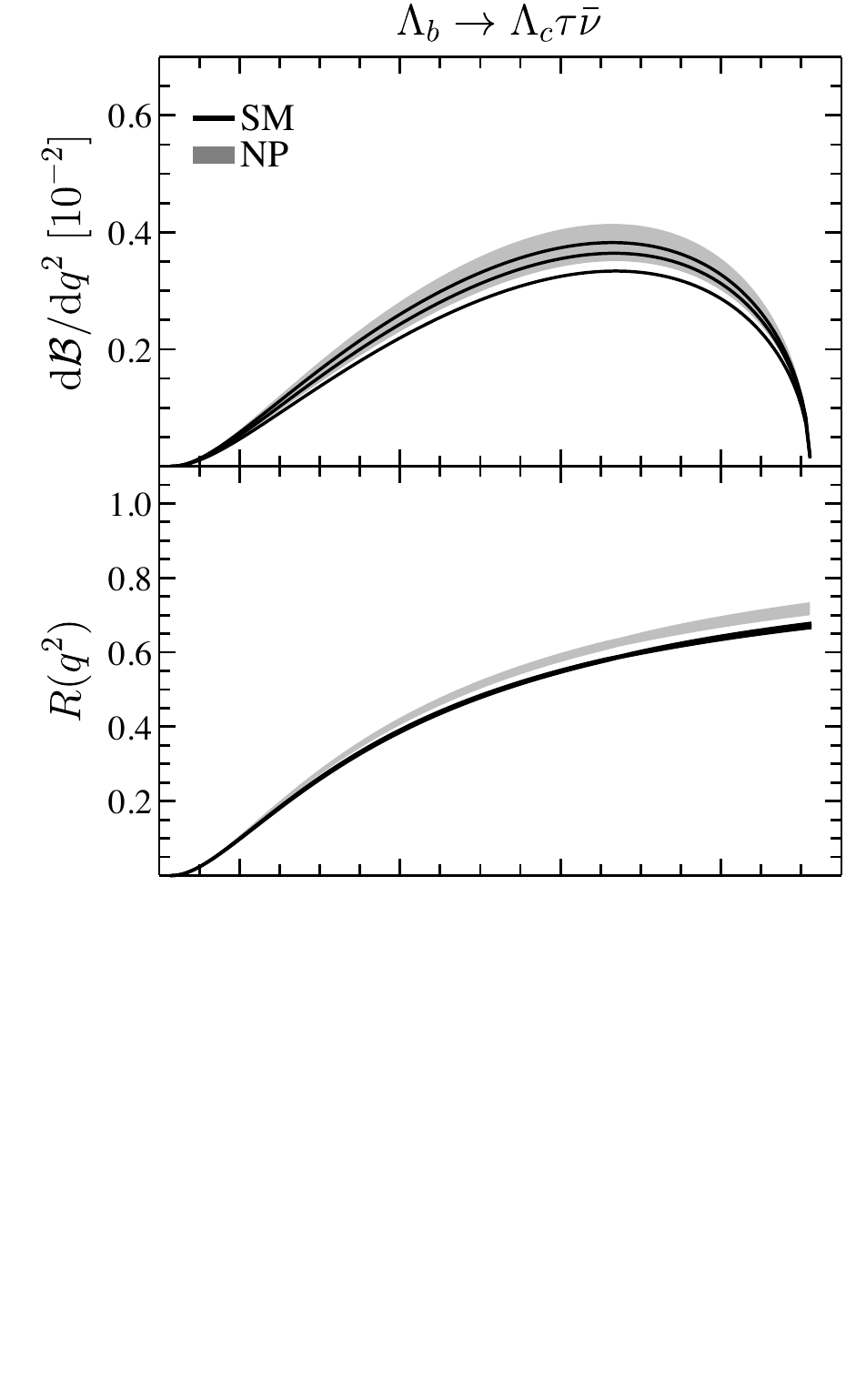}
  \qquad
  \includegraphics[width=0.4\textwidth]{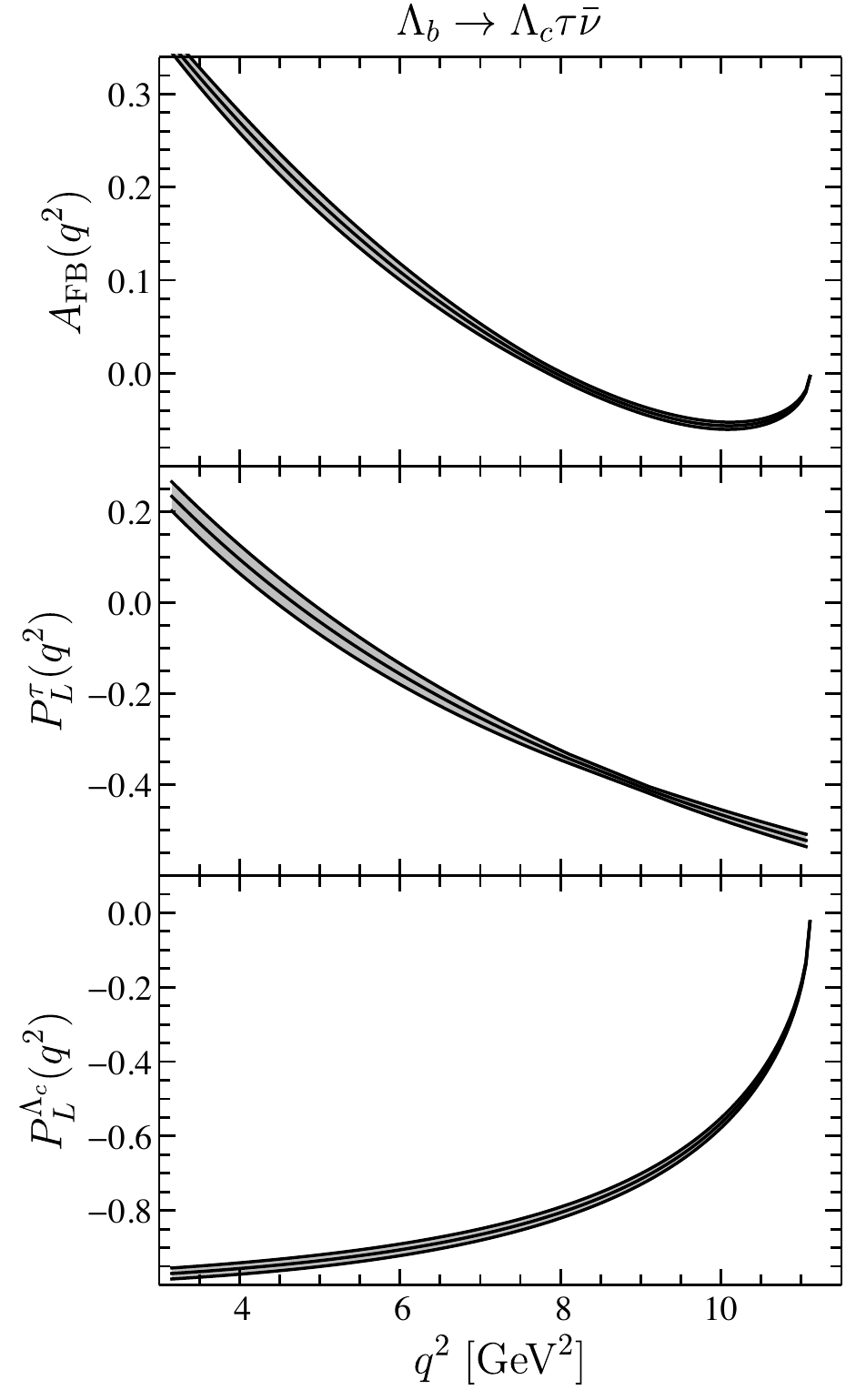}
  \caption{Differential observables in $\Lambda_b \to \Lambda_c \tau \bar\nu$ decay. Other captions are the same as in figure~\ref{fig:B}.}
  \label{fig:Lambdab}
\end{figure}

Using the allowed parameter space at $2\sigma$ level derived in the last subsection, we make predictions on the five $b \to c \tau \bar\nu$ decays, $B \to D^{(*)} \tau \bar\nu$, $B_c \to \eta_c \tau\bar\nu$, $B_c \to J/\psi \tau \bar\nu$, and $\Lambda_b \to \Lambda_c \tau \bar\nu$ decays. In table~\ref{tab:numerics}, the SM and SUSY predictions on the various observables in these decays are presented. The SUSY predictions have included the uncertainties induced by the form factors and CKM matrix elements. At present, there are no available measurements on the $B_c \to \eta_c \tau \bar\nu$ and $\Lambda_b \to \Lambda_c \tau \bar\nu$ decays. From table~\ref{tab:numerics}, it can be seen that, although the SUSY predictions for the branching fractions and the LFU ratios in these two decays overlap with their $1\sigma$ SM range, they can be considerably enhanced by the RPV SUSY effects.

Now we start to analyze the $q^2$ distributions of the differential branching fraction $\mathcal B$, the LFU ratio $R$, the lepton forward-backward asymmetry $A_{\rm FB}$, the polarization fraction of $\tau$ lepton $P_L^\tau$ and the polarization fraction of daughter meson ($P_{L}^{D^{*}}$, $P_{L}^{J/\psi}$, $P_L^{\Lambda_c}$). For the two ``$B \to P$'' transitions $B \to D \tau \bar\nu$ and $B_c \to \eta_c \tau \bar\nu$, their differential observables in the SM and RPV SUSY are shown in figure~\ref{fig:B}. It can be seen that all the differential distributions of these two decays are very similar, while the observables in $B_c \to \eta_c \tau \bar\nu$ suffer from larger theoretical uncertainties, which are due to large uncertainties induced by the $B_c \to \eta_c$ form factors. In the RPV SUSY, the branching fraction of $B \to D \tau \bar\nu$ decay can be largely enhanced, while the LFU ratio is almost indistinguishable from the SM prediction. Therefore, the differential distribution of $R_D(q^2)$ is hard to provide testable signature of the RPV SUSY. In addition, the RPV SUSY does not affect the forward-backward asymmetry $A_{\rm FB}$ and $\tau$ polarization fraction $P_L^\tau$ in these two decays, as shown in figure~\ref{fig:B}. The reason is that the RPV couplings only modify the Wilson coefficient $\mathcal C_{L,2}$ and its effects in the numerator and denominator in eqs.~(\ref{eq:AFB}) and (\ref{eq:PL}) are cancelled out exactly. This feature could be used to distinguish from the NP candidates, which can explain the $R_{D^{(*)}}$ anomaly but involves scalar or tensor interactions~\cite{Freytsis:2015qca,Bauer:2015knc,Li:2016pdv}.

The differential observables in the $B \to D^* \tau \bar\nu$ and $B_c \to J/\psi \tau \bar\nu$ decays are shown in figure~\ref{fig:Bc}. As expected, these two ``$B \to V$'' processes have very similar distributions. In these two decays, the enhancement by the RPV SUSY effects is not large enough to make the branching ratios deviate from the SM values by more than $1\sigma$. However, the LFU ratios $R_{D^*}(q^2)$ and $R_{J/\psi}(q^2)$ are significantly enhanced in the whole kinematical region, especially in the large dilepton invariant mass region. In this end-point region, the theoretical predictions suffer from very small uncertainties compared to the other kinematical region. By this virtue, the LFU ratios $R_{D^*}(q^2)$ and $R_{J/\psi}(q^2)$ in the RPV SUSY deviate from the SM predictions by about $2\sigma$ level. Therefore, future measurements on these differential ratios could provide more information about the $R_{D^{(*)}}$ anomaly and are important for the indirect searches for SUSY. In addition, as in the $B \to D \tau \bar\nu$ and $B_c \to \eta_c \tau \bar\nu$ decays, the angular observables $A_{\rm FB}$, $P_L^\tau$ and $P_L^{D^*,J/\psi}$ are not affected by the SUSY effects.

Figure~\ref{fig:Lambdab} shows the differential observables in the $\Lambda_b \to \Lambda_c \tau \nu$ decay. It can be seen that the RPV SUSY effects significantly enhance the branching fraction and the LFU ratio. Especially, at the large dilepton invariant mass, the ratio $R_{\Lambda_c}(q^2)$ in the SUSY shows about more than $2\sigma$ discrepancy from the SM values. With large $\Lambda_b$ samples at the future HL-LHC, this decay is expected provide complementary information to the direct SUSY searches. In addition, as in the other decays, the RPV SUSY effects vanish in the various angular observables.

\section{Conclusions}
\label{sec:conclusions}

Recently, several hints of lepton flavour universality violation have been observed in the experimental data of semi-leptonic $B$ decays. Motivated by the recent measurements of $P_L^{D^{*}}$, we have investigated the RPV SUSY effects in $b \to c \tau \bar\nu$ transitions. After considering various flavour processes, we obtain strong constraints on the RPV couplings, which are dominated by $\mB(B^+ \to \pi^+ \nu \bar\nu)$, $\mB(B^+ \to K^+ \nu \bar\nu)$, and $g_{Z\tau_L\tau_L}$. In the surviving parameter space, the $R_{D^{(*)}}$ anomaly can be explained at $2\sigma$ level, which results in bounds on the coupling products, $-0.082 < \lp1 \lps3 < 0.087$, $0.018 < \lp2 \lps3 < 0.057$, and $0.033  < \lp3 \lps3 < 0.928$. The upper bound on the coupling $\lp3$ prevents this coupling from developing a Landau pole below the GUT scale.

In the parameter space allowed by all the constraints, we make predictions for various flavour processes. For $B^+ \to K^+ \nu \bar\nu$ decay, a lower bound $\mB (B^+ \to K^+ \nu \bar\nu ) > 7.37 \times 10^{-6}$ is obtained. Compared to the SM prediction $(3.98 \pm 0.47) \times 10^{-6}$, this decay can provide an important probe of the RPV SUSY effects at Belle II. We also find interesting correlations among $R_D$, $R_{D^*}$, $\mB (B^+ \to K^+ \nu \bar\nu)$, $\mB(B \to \tau \nu)$, and $g_{Z\tau_L\tau_L}/g_{Z\ell_L\ell_L}$. For example, the RPV SUSY effects always enhance $\mB(B^+ \to K^+ \nu \bar\nu)$ and suppress $g_{Z\tau_L\tau_L}/g_{Z\ell_L\ell_L}$ simultaneously, which makes one of them must largely deviate from its SM value.

Furthermore, we have systematically investigated the RPV SUSY effects in five $b \to c \tau \bar\nu$ decays, including $B \to D^{(*)} \tau \bar\nu$, $B_c \to \eta_c \tau \bar\nu$, $B_c \to J/\psi \tau \bar\nu$ and $\Lambda_b \to \Lambda_c \tau \bar\nu$ decays, and focus on the $q^2$ distributions of the branching fractions, the LFU ratios, and various angular observables. It is found that the differential ratios $R_{D^*}(q^2)$, $R_{J/\psi}(q^2)$, and $R_{\Lambda_c}(q^2)$ are significantly enhanced by the RPV SUSY effects in the large dilepton invariant mass region. Although the integrated ratios $R_{D^*, J/\psi, \Lambda_c}$ in the SUSY overlap with $1\sigma$ range of the SM values, the differential ratios $R_{D^*, J/\psi, \Lambda_c}(q^2)$ in this kinematical region show about more than $2\sigma$ discrepancy between the SM and SUSY predictions. In addition, the SM and RPV SUSY predictions on the various angular observables are indistinguishable, since the RPV SUSY scenario does not generate new operators beyond the SM ones.

The decays $B^+ \to K^+ \nu \bar\nu$ and $B \to \tau \bar\nu$, as well as the differential observables in $b \to c \tau \bar\nu$ decays, have the potential to shed new light on the $R_{D^{(*)}}$ anomalies and may serve as a test of the RPV SUSY. With the forthcoming SuperKEKB and the future HL-LHC, our results are expected to provide more information on the $b \to c \tau \bar\nu$ transitions and could correlate with the direct searches for SUSY in the future high-energy colliders.

\section*{Acknowledgements}
We thank Jun-Kang He, Quan-Yi Hu, Xin-Qiang Li, Han Yan, Min-Di Zheng, and Xin Zhang for useful discussions. This work is supported by the National Natural Science Foundation of China under Grant Nos. 11775092, 11521064, 11435003, and 11805077. XY is also supported in part by the startup research funding from CCNU.

\begin{appendix}

\section{Form factors}
\label{sec:form factor}

For the operator in eq.~(\ref{eq:Heff}), the hadronic matrix elements of $B\to D$ transition can be parameterized in terms of form factors $F_+$ and $F_0$~\cite{Sakaki:2013bfa,Bardhan:2016uhr}. In the BGL parameterization, they can be written as expressions of $a_n^+$ and $a_n^0$~\cite{Bigi:2016mdz},
\begin{align}
  F_{+}(z) =\frac{1}{P_{+}(z)\phi_{+}(z,\mathcal{N})}\sum_{n=0}^{\infty} a_{n}^{+}z^n(w,\mathcal{N}),
  \qquad
  F_{0}(z) =\frac{1}{P_{0}(z)\phi_{0}(z,\mathcal{N})}\sum_{n=0}^{\infty} a_{n}^{0}z^n(w,\mathcal{N}),
\end{align}
where $z(w,\mathcal{N})=(\sqrt{1+w}-\sqrt{2\mathcal{N}})/(\sqrt{1+w}+\sqrt{2\mathcal{N}})$, $w=(m_B^2+m_D^2-q^{2})/(2m_Bm_D)$, $\mathcal{N}=(1+r)/(2\sqrt{r})$, and $r=m_D / m_B$. Values of the fit parameters    are taken from ref.~\cite{Bigi:2016mdz}.

For the $B \to D^*$ transition, the relevant form factors are $A_{0,1,2}$ and $V$. They can be written in terms of the BGL form factors as
\begin{align}
  A_{0}(q^2)& =\frac{m_{B}+m_{D^{*}}}{2\sqrt{m_{B}m_{D^{*}}}}P_{1}(w),
  \\
  A_{1}(q^2)& =\frac{f(w)}{m_{B}+m_{D^{*}}},\nonumber
  \\
  A_{2}(q^2)& =\frac{(m_{B}+m_{D^{*}})\left[(m_{B}^{2}-m_{D^{*}}^{2}-q^{2})f(w)-2m_{D^{*}}\mathcal{F}_{1}(w)\right]}{\lambda_{D^{*}}(q^{2})},\nonumber
  \\
  V(q^2)& =m_{B}m_{D^{*}}(m_{B}+m_{D^{*}})\frac{\sqrt{w^{2}-1}}{\sqrt{\lambda_{D^{*}}(q^{2})}}g(w),\nonumber
\end{align}
where $w=(m_B^2+m_{D^*}^2-q^2)/2m_Bm_{D^*}$ and $\lambda_{D^*}=[(m_B - m_{D^*})^2-q^2][(m_B+m_{D^*})^2-q^2]$. The four BGL form factors can be expanded as series in $z$
\begin{align}
  f(z) &= \frac{1}{P_{1+}(z) \phi_f(z)} \sum_{n=0}^\infty a_n^f z^n,
         & \mathcal F_1(z) &= \frac{1}{P_{1+}(z)\phi_{\mathcal F_1}(z)} \sum_{n=0}^\infty z_n^{\mathcal F_1} z^n,
  \\
  g(z) &= \frac{1}{P_{1-}(z)\phi_g(z)} \sum_{n=0}^\infty a_n^g z^n,
         & P_1(z) &= \frac{\sqrt r}{(1+r)B_{0-}(z) \phi_{P_1}(z)} \sum_{n=0}^\infty a_n^{P_1} z^n, \nonumber
\end{align}
where $z=(\sqrt{w+1}-\sqrt{2})/(\sqrt{w+1}+\sqrt{2})$ and $r=m_{D^*}/m_B$. Explicit expressions of the Blaschke factors $P_{1\pm}$ and $B_{0-}$ and  the outer functions $\phi_{i}(z)$ can be found in ref.~\cite{Bigi:2017njr,Bigi:2017jbd}. We also adopt the values of the fit parameters in ref.~\cite{Bigi:2017njr,Bigi:2017jbd}.

The $\Lambda_b\rightarrow\Lambda_c$ hadronic matrix elements can be written in terms of the helicity form factors $F_{0,+,\perp}$ and $G_{0,+,\perp}$~\cite{Detmold:2015aaa,Datta:2017aue}. Following ref.~\cite{Detmold:2015aaa}, the lattice calculations are fitted to two Bourrely-Caprini-Lellouch $z$-parameterization~\cite{Bourrely:2008za}. In the so called ``nominal fit'', a form factor has the following form
\begin{align}\label{eq:nominalfitphys}
f(q^2) = \frac{1}{1-q^2/(m_{\rm pole}^f)^2} \big[ a_0^f + a_1^f\:z^f(q^2)  \big], 
\end{align}
while the form factor in the ``higher-order fit'' is given by
\begin{align}\label{eq:HOfitphys}
f_{\rm HO}(q^2) =& \frac{1}{1-q^2/(m_{\rm pole}^f)^2} \bigl\lbrace a_{0,{\rm HO}}^f + a_{1,{\rm HO}}^f\:z^f(q^2) + a_{2,{\rm HO}}^f\:[z^f(q^2)]^2  \bigr\rbrace,
\end{align}
where $z^f(q^2) = (\sqrt{t_+^f-q^2}-\sqrt{t_+^f-t_0})/(\sqrt{t_+^f-q^2}+\sqrt{t_+^f-t_0})$, $t_0 = (m_{\Lambda_b} - m_{\Lambda_c})^2$, and $t_+^f = (m_{\rm pole}^f)^2$. Values of the fit parameters are taken from ref.~\cite{Datta:2017aue}.

In addition, the form factors for $B_c \to J/\psi $ and $B_c \to \eta_c$ transitions are taken from the results in the Covariant Light-Front Approach in ref.~\cite{Wang:2008xt}.

\end{appendix}

\end{document}